\documentclass[aps,prb,twocolumn]{revtex4-1}
\usepackage{graphicx,color}
\usepackage{hyperref}
\usepackage{dcolumn}
\usepackage{bm}
\usepackage{amsmath,amsthm,amssymb}
\usepackage{mathtools}

\begin{document}
\begin{flushright}
YITP-18-47
\end{flushright}
\title{
  Spin Connections for Nonrelativistic Electrons on Curves and Surfaces
}

\author{Toru Kikuchi}%
\affiliation{%
Yukawa Institute for Theoretical Physics,  Kyoto University, Kyoto 606-8502, Japan}

\begin{abstract}
We propose a basic theory of nonrelativistic spinful electrons on curves and surfaces.  In particular, we discuss the existence and effects of spin connections,  which describe how spinors and vectors couple to the geometry of curves and surfaces. 
We derive explicit expressions of spin connections  by performing  simple dimensional reductions from  three-dimensional flat space.   
The spin connections  act on electrons as spin-dependent magnetic fields, which are known as ``pseudomagnetic fields''  in the context of, for example, graphenes and Dirac/Weyl semimetals.  We propose that these spin-dependent magnetic fields are present universally on curves and surfaces, acting on electrons regardless of the nature of their spinorial degrees of freedom or their dispersion relations.   
We  discuss that, via the spin connections, the curvature effects will cause the spin Hall effect, and induce the Dzyaloshinskii--Moriya interactions between magnetic moments on curved surfaces, without relying on relativistic spin-orbit couplings.  We also note the importance of spin connections on the orbital physics of electrons in curved geometries.  
\end{abstract}

\date{\today}

\maketitle

Spinors are geometrical objects.  Spinors rotate, together with vectors, when we rotate a physical system \cite{book_Sakurai}.  They can also generate vectors through the familiar bilinear form $\psi^\dagger \sigma^i\psi$ using the Pauli matrices $\sigma^i$.   Being geometrical objects, spinors can couple to the  background geometry in which they live.  Spin connections  
\cite{book_Carroll, book_Nakahara, Kamien2002, book_David, book_GSW,  book_ParkerToms} describe how spinors (and vectors) couple to their background geometry.  
Spin connections have been used to study spinors in curved spacetimes in the realm of general relativity.  The same formalism can also be applied  to nonrelativistic spinors on curves and surfaces embedded in our daily three-dimensional flat space.  

What spin connections are, and why they are (expected to be) present on curves and surfaces, can be understood as follows.  For comparison, let us first consider the case where we rigidly rotate a  bulk material at an angular velocity $\bm \Omega$.  In the rotating material, the physics of an electron is dragged by the rotation of its surrounding environment (e.g., a lattice).  Such dragging arises from  quantum mechanical interactions owing to the overlap between the wavefunctions of the electron and the environment.   As a result,  the magnetic moment $\bm M$ of an electron, for example, is forced to rotate together with  its environment; its time derivative (for a laboratory observer) changes as $\partial_t\bm M\rightarrow \partial_t \bm M-\bm \Omega\times\bm M$.  
This is known as the Barnett effect \cite{Barnett1915, Barnett1935}.    What happens to electrons on curves and surfaces can be regarded as the spatial counterpart of the Barnett effect.  On curves and curved surfaces, the local environment of an electron gradually rotates as an electron propagates in the geometry, where the local anisotropy of the system is characterized by the tangent vectors $\bm T$ for curves and the normal vectors $\bm n$ for surfaces.  The local physics of an electron is dragged by the rotation of its local environment as it propagates.  The situation is rather analogous to the Barnett effect;  the only difference is that spatial derivatives are involved rather than time derivatives.  In terms of the magnetic moment, the spatial derivative $\partial_\mu$ tangential to a curve or surface changes as $\partial_\mu\bm M \rightarrow \partial_\mu\bm M-\bm \Omega_\mu \times \bm M$, with $\bm \Omega_\mu\sim \partial_\mu \bm T$ for curves and $\bm \Omega_\mu\sim \partial_\mu \bm n$ for surfaces.  This $\bm \Omega_\mu$ is to be called the spin connection on curves and surfaces \cite{book_David, Kamien2002}.  Because the magnetic moment is related to a spinor $\psi$ as $\bm M=\psi^\dagger \bm \sigma \psi$ with the Pauli matrices $\sigma^i$, a spinor  also rotates as it propagates on a curve or surface; its spatial derivative changes as $\partial_\mu\psi\rightarrow (\partial_\mu+i\bm \Omega_\mu\cdot \frac{\bm \sigma}{2})\psi$.  In this way, electrons couple to the geometry in which they exist, and the spin connections $\bm \Omega_\mu$ describe the coupling.

In the case of curves, the discussion so far can be rephrased as follows.  Let us regard a curve as being deformed from a straight line.  Each infinitesimal portion of the curve is related to the original portion of the straight line by a rotation.  Then, using the corresponding SU(2) rotation matrix $\mathcal U$, the Hamiltonian density $\mathcal H_{\rm curve}$ of the portion of the curve is related to the Hamiltonian density $\mathcal H_{\rm line}$ of the straight line by  $\mathcal H_{\rm curve}=\mathcal U^\dagger \mathcal H_{\rm line} \mathcal U$ (cf. Ref.\cite{Stockhofe2014}).  In particular, the derivative operator $\partial_\mu \psi$ is transformed as $\mathcal U^\dagger \partial_\mu (\mathcal U\psi)=(\partial_\mu+i\bm \Omega_\mu \cdot \frac{\bm \sigma}{2})\psi$ with  $i\bm \Omega_\mu \cdot \frac{\bm \sigma}{2}=\mathcal U^\dagger\partial_\mu \mathcal U$.  Thus, the spin connection $\bm \Omega_\mu$ appears due to the position-dependent rotation of each portion of the curve with respect to the straight line.  In the case of surfaces, the discussion is more complicated due to the intrinsic curvature of surfaces, and the derivation of spin connections should be performed more systematically as will be described in this paper.  

We also note the necessity of spin connections in view of the Dirac theory.   
Spin connections are naturally required in order for the Dirac theory on a curve or surface to be Hermitian or to yield the Schr\"{o}dinger equation as its nonrelativistic limit  (see Appendix \ref{app: Dirac} for details).  
A related discussion is given by Meijer {\it et. al.}\cite{Meijer2002}.  They discussed that the relativistic spin-orbit coupling (SOC) term on a curved geometry needs a correction;   without this correction, the SOC term is not  Hermitian.  Interestingly, their correction coincides with the spin connection  to be derived in this paper [Eq.\eqref{spin connection for curves}].  Meijer {\it et. al.} added the correction only to the SOC term,  not  to the usual kinetic term.  However, because both the kinetic term and the SOC term have the Dirac theory as the common origin, these terms should be treated in a unified way;  the correction, i.e., the spin connection, should be added also to the usual kinetic term.  

In most previous work in condensed matter physics that studies nonrelativistic spinful electrons on curves and surfaces, spin connections have not been taken into account.  (An exception is the study of the quantum Hall effect \cite{Wen1992, Froehlich1993}, where spin connections acting on the ``orbital spin'' are introduced.)   There seem to be mainly two reasons for this situation.  

The first reason is that the so-called ``thin-layer approach'' \cite{Jensen1971, DaCosta1981}, originally used for the spinless case, has been applied too straightforwardly to the spinful case.    
In this approach, for the case of surfaces,  one starts from a three-dimensional bulk Hamiltonian  
and reduces the thickness of the system,  to derive a surface Hamiltonian.  It is difficult to apply this approach to the spinful case.   
Physically, an electronic state near a surface is affected by the surface normal direction $\bm n$; both a spinor and the normal direction $\bm n$ are directional quantities and they are generally coupled.  
However, a three-dimensional bulk Hamiltonian does not contain such coupling terms, simply because it describes only a bulk and ignores the terms which are present only near the surface.  
It is therefore difficult to derive a surface Hamiltonian from such a bulk Hamiltonian with no information about the surface, only  by narrowing the domain of the bulk Hamiltonian in the thin-layer approach.

The second reason 
is that the previous studies where spin connections for electrons on curves and surfaces are taken into account have mainly focused on  the case of (quasi-)relativistic Dirac dispersion relation \cite{Gonzalez1992, Kane1997, Lee2009, Zhang2010}. 
This may be partly because the formalism of general relativity  applies directly to this case; or partly because the Dirac dispersion relation appears in graphene, the most representative and mechanically flexible two-dimensional material.  
However, at least in principle, there is no reason for  spin connections to be  relevant only to a specific dispersion relation.

In this paper, we introduce the concept of spin connections and derive their expressions on curves and surfaces in as simple a way as possible. 
Then, we discuss the basic properties of nonrelativistic electrons constrained on the geometries.   

Let us first consider the case of surfaces.  We will always make  summations over repeated indices.  
Let us set a curvilinear coordinate system $x^M$ in three-dimensional flat space so that,   when we divide the index into $M=(\mu,\perp)$, the coordinate $x^\perp$ is in the direction normal to the surface, 
and $x^\mu$ ($\mu=1,2$) parametrizes the surface.    
We also use the indices $i,j,\dots = x,y,z$ to describe  Cartesian coordinates $x^i$.  The transformation matrices between these coordinates are $e^M_i\equiv \partial x^M/\partial x^i$ and its inverse $e^i_M \equiv \partial x^i/\partial x^M$.
When a vector field $V^i(x^\mu)$ on the surface is measured in  Cartesian coordinates, the usual differentiation, which we call here  the ``flat differentiation'' $\nabla^{({\rm flat})}$, acts on  $V^i$ simply as $\nabla^{({\rm flat})}_\mu V^i=\partial_\mu V^i$.  
On the other hand, when a vector is measured in the curvilinear coordinates as $V^M=e^M_i V^i$, then $\nabla^{({\rm flat})}$ acts as  $\nabla^{({\rm flat})}_\mu V^N = \partial_\mu V^N + \Gamma^{N}_{~\mu L}V^L$.   
The quantity $\Gamma^N_{~\mu L}$ is called a ``connection'' \cite{book_FosterNightingale, book_Poisson, book_SatoKatsu, annotation_indices};  it is determined by the relation $\nabla^{({\rm flat})}_\mu V^N=e^N_i\nabla^{({\rm flat})}_\mu V^i$, and is given by $\Gamma^N_{~\mu L}= e^{N}_j\partial_\mu e^j_L$.

Compared with $\nabla^{({\rm flat})}$, we can define another kind of differentiation  with a nontrivially curved connection, which is determined by the geometry of the surface.  We call it here the ``curved derivative'', and express it as $\nabla$.   
The largest difference between these two types of differentiation, $\nabla^{({\rm flat})}$ and $\nabla$, is the way they act on the normal component $V^\perp$ of a vector $V^M$.  A two-dimensional observer living on the surface (e.g., an electron) has the freedom to change his coordinate system $x^\mu$ on the surface to a new one, $x^\mu\rightarrow x^{\mu '}$, but he observes that $V^\perp$ is invariant under this coordinate transformation.  
Then, $V^\perp$ is a scalar for him, whereas, for us, it is just a particular component of the three-component vector $V^M$.  
The curved derivative $\nabla$ takes the viewpoint of this two-dimensional observer and acts as $\nabla_\mu V^\perp=\partial_\mu V^\perp$ since $V^\perp$ is just a scalar. On the other hand, the flat derivative acts as $\nabla^{({\rm flat})}_\mu V^\perp=\partial_\mu V^\perp + \Gamma^\perp_{~\mu N}V^N$.  This means that $\Gamma^\perp_{~\mu N}$ is truncated to be zero for the curved derivative $\nabla_\mu$.   
The same reasoning for $V_\perp$ requires that $\Gamma^N_{~\mu \perp}$ is also truncated for $\nabla_\mu$.  These truncations also determine how $\nabla_\mu$ acts on the tangential components $V^\nu$ of a vector $V^M$: $\nabla_\mu V^\nu=\partial_\mu V^\nu + \Gamma^\nu_{~\mu\lambda}V^\lambda$.  
To summarize, the flat derivative $\nabla^{({\rm flat})}_\mu$ and the curved derivative $\nabla_\mu$ act differently on the curvilinear indices as
\begin{align}
&\nabla^{({\rm flat})}_\mu V^N = \partial_\mu V^N + \Gamma^{N}_{~\mu L}V^L;
\notag \\
&\nabla_\mu V^\perp = \partial_\mu V^\perp, ~~
\nabla_\mu V^\nu = \partial_\mu V^\nu + \Gamma^{\nu}_{~\mu\lambda}V^\lambda.
\label{difference curvilinear}
\end{align}
After all, dimensional reduction of $\Gamma^M_{~NL}$ is performed for $\nabla$, where the direction to be reduced is the normal direction  at each point of the surface.  

\begin{figure}
	\begin{center}
		\includegraphics[scale=0.4]{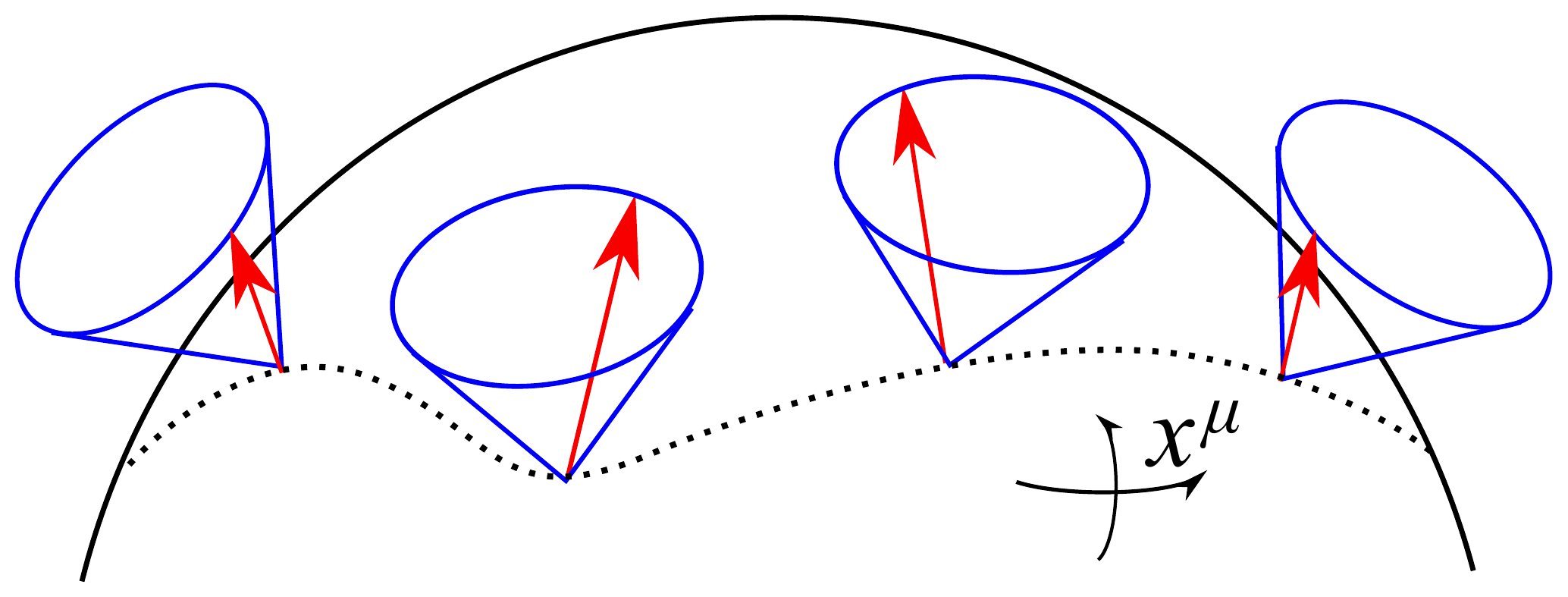}
		\caption{
		    Schematic illustration of the transport of a vector on a surface.  A vector $V^i$ (red arrow) is transported by $\nabla_\mu \bm V=\partial_\mu \bm V - \bm \Omega_\mu \times \bm V=0$ with the spin connection $\bm \Omega_\mu$ along an arbitrary trajectory (dashed curve) on a surface.  The component of the vector normal to the surface remains constant, whereas the tangential components are generally rotated in the tangential planes (unless the transport trajectory coincides with a geodesic of the surface).  As a result, the vector undergoes  precession around the normal vectors of the surface as it is transported (in the figure, the precession rate is drawn somewhat exaggeratedly).   The blue cones are the precession cones, in which the direction connecting the apex and the center of the base circle coincides with the normal direction at each point of the surface.  
			}
			\label{fig:Fig1}
	\end{center}
\end{figure}

Next, let us look at the way the curved derivative $\nabla$ acts on a vector  $V^i$ with Cartesian indices.  It is given by $\nabla_\mu V^i=e_M^i\nabla_\mu V^M$, which becomes $\nabla_\mu V^i = \partial_\mu V^i + \Omega_\mu^{ij}V^j$, with
\begin{equation}
\Omega_\mu^{ij} = e^{i}_{\nu} \Gamma^\nu_{~\mu\lambda} e^{\lambda}_j 
+ e^{i}_{N}\partial_\mu e^N_j.
\label{2d transformation}
\end{equation}
We call $\Omega_\mu^{ij}$ the ``spin connection'' on a surface: it is a connection represented in Cartesian indices \cite{annotation_specific}.  
We can express $\Omega_\mu^{ij}$ as follows in  terms of the surface normal vector $n^i=\partial x^i/\partial x^\perp$ at each point on the surface.  By subtracting from Eq.\eqref{2d transformation} the relation $0 = e^{i}_{N} \Gamma^N_{~\mu L} e^{L}_j 
+ e^{i}_N\partial_\mu e^N_j$, which follows immediately from $\Gamma^N_{~\mu L}=e^N_i\partial_\mu e_L^i$, we can see that the spin connection is given by the truncated components of the connection $\Gamma^M_{~NL}$  (see Appendix \ref{app: eq3} for details):
\begin{align}
\Omega_\mu^{ij} =& - e^i_\perp \Gamma^\perp_{~\mu\lambda}e^{\lambda}_j - e^i_\nu \Gamma^\nu_{~\mu\perp}e^{\perp}_j
\notag \\
=& n^i\partial_\mu n^j - n^j\partial_\mu n^i.
\label{Omegaij expression}
\end{align}
As seen explicitly in Eq.\eqref{Omegaij expression}, $\Omega_\mu^{ij}$ is antisymmetric with respect to $i$ and $j$, and it can therefore be written as $\Omega_\mu^{ij}=\varepsilon_{ijk}\Omega_\mu^k$.  
This means that, when a vector is transported by $\nabla_\mu V^i= (\partial_\mu \bm V - \bm \Omega_\mu\times \bm V)^i=0$, the vector spins (or precesses) at the rate $\bm \Omega_\mu$, as in Fig.1.      
Hence its name, a {\it spin} connection, is natural.   
In particular, $\nabla_\mu n^i=0$: the spin connection expresses the way the normal vector rotates as we move on the surface.  
To summarize, the flat and curved derivatives, $\nabla_\mu^{({\rm flat})}$ and $\nabla_\mu$, act 
differently on the Cartesian indices:   
\begin{equation}
\nabla_\mu^{({\rm flat})}V^i=\partial_\mu V^i;~~~\nabla_\mu V^i = \partial_\mu V^i - \varepsilon_{ijk}\Omega_\mu^j V^k,
\label{difference Cartesian}
\end{equation}
which is equivalent to Eq.\eqref{difference curvilinear}.

We have so far discussed vectors.  Let us next consider spinors and the way the curved derivative $\nabla$  acts on them.   
We can easily see this by taking the spin magnetic moment vector $V^i=\psi^\dagger \sigma^i\psi$ with $\psi$ a spinor and $\sigma^i$ the Pauli matrices.    
For the curved derivative $\nabla$ to act on the magnetic moment as in Eq.\eqref{difference Cartesian},  
the derivative must act on a spinor as 
\begin{equation}
\nabla_\mu \psi = \left(\partial_\mu + i\bm \Omega_\mu\cdot \frac{\bm\sigma}{2}\right)\psi~~~{\rm with}~~\bm \Omega_\mu = \bm n\times \partial_\mu \bm n, 
\label{covariant derivative}
\end{equation}
assuming the Leibniz rule for the derivative.  
 The expression of the spin connection  in Eq.\eqref{covariant derivative} agrees with those in, for example, Refs.\cite{Kavalov1986, Zhang2010, annotation_David}.  
As a spinor propagates on a surface, it  receives a torque and gets rotated at the rate $\bm \Omega_\mu$.  Physically, this torque is to be interpreted as being exerted by the environment (e.g., a lattice), which confines the spinor to the surface.  The normal vector $\bm n$ gives the local anisotropy of the environment, and $\bm \Omega_\mu\sim \partial_\mu \bm n$ is  the rotation rate of such local anisotropy.

Let us next study some of the basic properties of electrons confined on surfaces.    
We can see vividly the effects of the spin connection by rotating the spinor frame to what we call here the ``intrinsic frame''.  We have so far used the laboratory spinor frame, where we take the spin-quantization axis to lie along the $z$ direction in the laboratory (Cartesian) coordinates.  
For electrons on a surface, a more natural frame is the intrinsic spinor frame where the spin-quantization axes are taken to lie along the directions of the normal vectors $\bm n$ distributed over the surface.
The transformation from the laboratory spinor frame $\psi$ to the intrinsic spinor frame $\widetilde\psi$ is described by $\psi=U\widetilde\psi$ with an SU(2)  matrix $U$ satisfying $U^\dagger \bm n\cdot \bm \sigma U=\sigma^3$, where $\sigma^3=\begin{psmallmatrix}1& 0 \\ 0 & -1\end{psmallmatrix}$.  The spin connection $\omega_\mu$ in the intrinsic spinor frame is defined as $(\partial_\mu+i\Omega_\mu)\psi =U(\partial_\mu+i\omega_\mu)\widetilde\psi$, where $\Omega_\mu\equiv \bm \Omega_\mu\cdot \frac{\bm \sigma}{2}$, and it is given by (see Appendix \ref{app: eq5} for details)
\begin{align}
\omega_\mu =& U^\dagger \Omega_\mu U - iU^\dagger \partial_\mu U
\notag \\
=& (1-\cos\theta)\partial_\mu \phi ~\frac{\sigma^3}{2},
\label{intrinsic spin connection}
\end{align}
with $n^i=(\sin\theta\cos\phi, \sin\theta\sin\phi,\cos\theta)^i$.  
From Eq.\eqref{intrinsic spin connection}, we see that the intrinsic frame diagonalizes the spin connection.  
In particular, the Schr\"{o}dinger equation for a freely propagating electron reads $i\hbar \partial \widetilde\psi/\partial t = -(\hbar^2/2m_{\rm e})\mathcal D^2\widetilde\psi$ with $m_{\rm e}$ being the electron mass,  and the Laplacian $\mathcal D^2$ is given by $\mathcal D^2\widetilde\psi \equiv g^{\mu\nu}\left(\nabla_\mu \nabla_\nu  - \Gamma^\lambda_{~\mu\nu}\nabla_\lambda \right)\widetilde\psi =\frac{1}{\sqrt{g}}\nabla_\mu\left(\sqrt{g}g^{\mu\nu}\nabla_\nu\right)\widetilde\psi$.  
Here, $g_{\mu\nu}\equiv (\partial x^i/\partial x^\mu)(\partial x^i/\partial x^\nu)$ is the induced metric on the surface and $\nabla_\mu$ is given as in Eq.\eqref{covariant derivative} with $\Omega_\mu$ replaced by $\omega_\mu$.   The spin-up and down components decouple in the Schr\"{o}dinger equation; an electron with  spin in the $\pm \bm n$ direction propagates keeping the $\bm n$-up/down states.  

The expression in Eq.\eqref{intrinsic spin connection} is that of the gauge field of a magnetic monopole \cite{book_Nakahara}.  Monopole gauge fields with magnetic charges $\pm\frac{1}{2}$ are coupled with the spin $\bm n$-up/down components.  As a result, the spin $\bm n$-up  and  $\bm n$-down components are subjected to the `magnetic fields' $B_{\rm m}\bm n$ and $-B_{\rm m}\bm n$, respectively (see Appendix \ref{app: pseudomagnetic field} for details).  The field strength $B_{\rm m}$ is given by $B_{\rm m}=K/2$, where $K\equiv{\rm det}(\nabla^{({\rm flat})}_\mu n^\nu)$ is the Gaussian curvature \cite{Kamien2002, book_David, book_doCarmo} at a point on the surface (e.g., $K=a^{-2}$ everywhere on a sphere of radius $a$).  When $K\sim 1~{\rm nm}^{-2}$, then $(\hbar/e)B_{\rm m}\sim 3\times10^2~{\rm tesla}$, with $e$ the electric charge.
Owing to this field strength of the spin connection, the curvature effects cause  cyclotron motions of electrons, which are in  opposite directions for the spin $\bm n$-up and $\bm n$-down components, yielding a vortical spin current circulating around each point of the surface.   

This kind of spin-dependent gauge field is known as a strain-induced \cite{annotation_strain} ``pseudomagnetic field'' in, for example,  graphenes \cite{Gonzalez1992, Neto2009, Levy2010, Vozmediano2010, book_Katsnelson, Amorim2016, Yi2017, Naumis2017}, transition-metal dichalcogenides \cite{Cazalilla2014} and Dirac/Weyl semimetals 
\cite{Cortijo2015} (see also Ref.\cite{Shapourian2015}). 
 The  main interest seems to have been focused on pseudomagnetic fields acting on valley or orbital degrees of freedom rather than on the true spin (exceptions are, e.g., Refs\cite{Lee2009, Zhang2010}, which treat spin connections acting on electron spin). Interest has also been focused on the case of the Dirac/Weyl dispersion relations.   
As discussed at the beginning of this paper, spin connections effectively represent the effects of the local rotation of the environment on electronic quantum states;   their presence is quite general,  regardless of the kinds of spinorial degrees of freedom (whether spin or pseudospin) \cite{annotation_nongeometric_indices} and the electron dispersion relations.   

The curvature effects are expected to induce the spin Hall effect, 
because the spin-up and -down components of electrons undergo $\bm E\times (\pm\bm B_{\rm m})$ drifts in opposite directions, where $\bm E$ is an externally applied electric field and $\pm\bm B_{\rm m}$ are the  pseudomagnetic fields.  
We also observe that electrons on surfaces generally have finite spin currents in their ground states:  
because  the surface curvature $K$ is generally position dependent,  the strengths of vortical spin currents caused by the pseudomagnetic fields $\pm (K/2)\bm n$ are also position dependent, and  the vortical spin currents are not canceled among nearby points; the net result is a finite spin current flowing on the surface.   
In a flat space, spin currents in ground states have been known to be induced by  relativistic spin-orbit coupling with broken inversion symmetry (such as the Rashba effect) \cite{Rashba2003}.  It has recently been shown \cite{Katsnelson2010, Kikuchi2016, Koretsune2018} that spin currents are a direct origin of the Dzyaloshinskii--Moriya interaction \cite{book_YosidaEnglish, book_Skomski, book_Kanamori, book_Yosida} in magnets.   
Therefore, curvature-induced spin currents in ground states on surfaces will lead to  intricate magnetic interactions, which we can engineer by the geometry of  surfaces.  
More details  will be reported elsewhere.


\begin{figure}
	\begin{center}
		\includegraphics[scale=0.5]{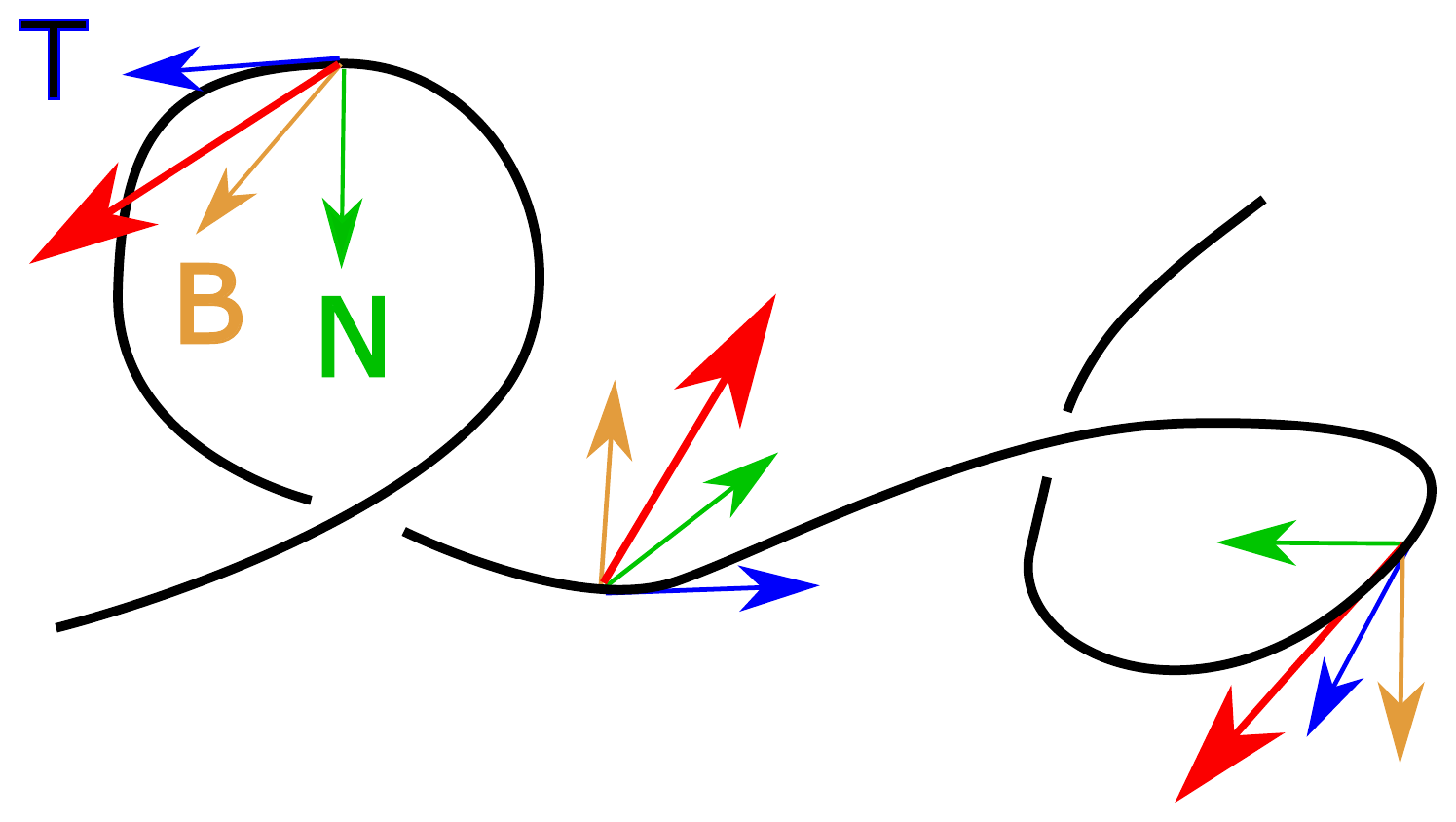}
		\caption{
			Schematic illustration of the transport of a vector along a curve.  A vector $V^i$ (red arrow) is transported by $\nabla_s \bm V = (d/ds)\bm V - \bm \Omega_s \times \bm V=0$, where $s$ is  the arc length parametrizing the curve.  The spin connection $\bm \Omega_s$ is given by the rotation rate of the Frenet--Serret frame $(\bm T, \bm N, \bm B)$ (represented by the blue, green and yellow arrows, respectively) along the curve.  The vector $\bm V$ rotates together with the Frenet--Serret frame as it is transported: its length and the   relative angles with respect to the Frenet--Serret frame remain constant during transport.  
			}
			\label{fig:Fig2}
	\end{center}
\end{figure}

Next, let us consider spin connections on curves, which 
 seem to have been discussed rarely in the literature.   
For a curve, a natural orthonormal frame \cite{book_doCarmo} is defined at each point  by the unit vectors $\bm T(s)\equiv \bm \gamma'(s)$, $\bm N(s)\equiv \bm T'(s)/|\bm T' (s)|$, and $\bm B(s) \equiv \bm T(s)\times \bm N(s)$, where $\gamma^i(s)$ specifies a point on the curve and  $s$ is the arc length parametrizing the curve.  We define a curvilinear coordinate system $x^M$ with $M=s,n,b$ by $x^i(x^M)=\gamma^i+ x^n N^i + x^b B^i$.  
Following the same procedure as in the case of surfaces, we truncate all  the components of $\Gamma^M_{~NL}$ except the tangential component $\Gamma^s_{~ss}$.  Then, the spin connection for a curve is given by (see Appendix \ref{app: eq7} for details) 
\begin{equation}
\bm \Omega_s = \kappa \bm B + \tau \bm T
\label{spin connection for curves}
\end{equation}
in the laboratory frame, where $\kappa\equiv \bm N\cdot \bm T'$ and $\tau \equiv \bm B\cdot \bm N'$ are the curvature and the torsion of the curve, respectively.  The quantity $\bm \Omega_s$ in Eq.\eqref{spin connection for curves} satisfies $d \bm X/ds = \bm \Omega_s\times \bm X$ for $\bm X=\bm T, \bm N, \bm B$ (known as the Frenet--Serret formulas \cite{book_Matsuda, book_doCarmo}), representing the way the orthonormal frame rotates as it moves along the curve. The nontrivially curved derivative for a spinor on a curve reads $\nabla_s\psi\equiv(\frac{d}{ds} + i\bm \Omega_s\cdot \frac{\bm \sigma}{2})\psi$.  A vector or a spinor transported by $\nabla_sV^i=0$ or $\nabla_s\psi=0$, respectively, changes its direction following the rotation of the $(\bm T, \bm N, \bm B)$ frame, as in Fig. 2.   When we go from the laboratory frame to the intrinsic frame consisting of $(\bm T, \bm N, \bm B)$ as its basis vectors,  the spin connection vanishes (see Appendix \ref{app: gauge trf for curves}).  This is  expected, because a one-dimensional curve is geometrically trivial in the intrinsic sense: its internal metric is always $g_{ss}=1$.  

Finally, let us mention that spin connections also act on atomic orbitals ($p$-orbitals, $d$-orbitals, etc.) of electrons,  where the orbital degrees of freedom are expressed as an additional index $\alpha$
on a wavefunction $\psi_\alpha$.  
The discussion is completely parallel to the spin-$\frac{1}{2}$ case.  The nontrivial derivative acting on the orbital magnetic moment $\psi^\dagger \ell^i\psi$, with the orbital angular momentum matrix $\bm \ell$, implies that it also acts on a spinor as $\nabla_\mu\psi=(\partial_\mu + i\bm \Omega_\mu \cdot \bm \ell)\psi$. 
The explicit forms of $\ell^i$ are   
 given as $\ell^i_{\alpha\beta}=-i\varepsilon_{i\alpha\beta}$ for $p$-orbitals,   $\ell^i_{\alpha\beta}=-2i\varepsilon_{ijk}\left([\xi^\alpha,\xi^\beta]\right)_{jk}$ for $d$-orbitals \cite{Coury2016}, etc. 
The spin connections acting on the orbitals will be useful  for describing the orbital physics on curved geometries   
made of, for example, transition-metal dichalcogenides \cite{Wang2012, Choi2017, Qin2017, Manzeli2017, Dong2017}.    Similar to the spin-$\frac{1}{2}$ case, the orbital Hall effect \cite{Bernevig2005, Kontani2008PRL, Tanaka2008, Kontani2009PRL}, and orbital angular-momentum currents in  ground states, will be induced purely by the curvature effects.  More generically, spin connections act on all kinds of particles (e.g., atoms, 
 magnons 
  and photons), 
which have their own angular momenta.

In summary, we have discussed the physics of spin connections on curves and surfaces, derived their explicit expressions with their geometrical meaning clarified, and studied their basic effects on electrons.   
Many fundamental questions of electrons on curves and surfaces are to be investigated.  Derivation of spin connections from more microscopic viewpoints should be performed for their theoretical foundation.  Theories on 
transportation phenomena \cite{Bernevig2006, Matsuo2011, Matsuo2013, Cazalilla2014, Qian2014, Matsuo2017, Kobayashi2017} and order-parameter physics 
\cite{Gaididei2014, Pylypovskyi2015, Sheka2015JPMT, Streubel2016,  Otalora2016,  Tretiakov2017, Charilaou2017, Moreno2017, Ball2017, Vojkovic2017} 
on curved geometry are necessary to be compared with experiments.  The presence of spin connections on curved geometry will enrich many classic physics such as the Hubbard model \cite{Zvyagin2017}, the Kondo effect \cite{Kondo2005},  superconductivity \cite{Maeno2012, Shekhter2016}, and the quantum Hall effect \cite{Can2014, Can2016},  which are usually studied on flat geometry.  Photonic crystals and possibly optical lattices give a highly controllable platform for curvature physics \cite{Szameit2010, Schultheiss2010}.  Spintronics using circuits  is technologically important application \cite{Nagasawa2012, Nagasawa2013, Saarikoski2015, Saarikoski2016, Avishai2017}.  
Spin connections arising from phonon excitations and lattice deformation \cite{Hamada2015, Dong2018} are essential ingredients to fully understand mechanical effects on electrons in solids, where the backreaction on lattice will  also have intriguing effects \cite{Brendel2017}.  
It is  of interdisciplinary interest to study biochemical objects such as DNA helices from physics viewpoints \cite{Gohler2011, Guo2012, Gutierrez2013}.  
 
\acknowledgements
The author thanks Y. Avishai, T. Kawakami, M. Sato, A. Shitade,  K. Taguchi, R. Takashima, K. Totsuka and P. Wiegmann for helpful comments.  In particular, the author thanks H. Saarikoski for comments and reading the manuscript, and  M. Matsuo for his generosity:  when the author discussed with him, it turned out that he had long had a similar idea independently.  The author is a Yukawa Research Fellow supported by the  Yukawa Memorial Foundation.  


\appendix

\section{
The Dirac theory favors nontrivial spin connections.}\label{app: Dirac}

Here, we show that the relativistic Dirac theory on curves and surfaces imposes the presence of nontrivial spin connections for its self-consistency, and, as the result, show that non-relativistic theories obtained as descendants of the Dirac theory  should include nontrivial spin connections.      

Let us first consider a surface.  We employ the curvilinear coordinate system $x^M=(x^\mu,x^\perp)$ and the Cartesian coordinate system $x^i=(x,y,z)$ as in the main text.  The full Dirac theory needs a four-component Dirac spinor $\Psi$.  
The Dirac action on a surface embedded in a four-dimensional flat Minkowski spacetime will read
\begin{equation}
S = -i\int dtd^2x\sqrt{g}\Psi^\dagger \gamma^0
\left[\gamma^0\partial_t + e^\mu_i\gamma^i(\partial_\mu+i\Omega_\mu) + m_{\rm e}\right]\Psi,
\label{Dirac Hamiltonian}
\end{equation}
where $e_i^\mu\equiv \partial x^\mu/\partial x^i$ and $g_{\mu\nu}\equiv e^i_\mu e^i_\nu$ with $e^i_\mu\equiv \partial x^i/\partial x^\mu$ is the induced metric  on the surface,  and $m_{\rm e}$ is the electron mass. 
We employ the natural unit, $\hbar=1$ and $c=1$.   The $\gamma$-matrices are given by  $\gamma^0=-i\begin{psmallmatrix}1& 0 \\ 0 & -1\end{psmallmatrix}$ and $\gamma^i=-i\begin{psmallmatrix}0& \sigma^i \\ -\sigma^i & 0\end{psmallmatrix}$ in the Dirac representation, satisfying $(\gamma^0)^2=-1$, $\{\gamma^i,\gamma^j\}=2\delta^{ij}$ and $\{\gamma^0,\gamma^i\}=0$ [we take the Minkowski metric as ${\rm diag}(-1,1,1,1)$].  We assume that the spin connection $\Omega_\mu$ has an expansion as $\Omega_\mu=\frac{1}{2}\Omega_\mu^{ij}\Sigma^{ij}$ with $\Sigma^{ij}\equiv -\frac{i}{4}[\gamma^i,\gamma^j]$.

Even if we set the spin connection $\Omega_\mu^{ij}$ undetermined when writing down Eq.\eqref{Dirac Hamiltonian},  the Hermitian (reality) condition of the action, $S^\dagger=S$, imposes a condition on $\Omega_\mu^{ij}$ as 
\begin{equation}
g^{\mu\nu}(\mathfrak{D}_\mu e_\nu^i +\Omega_\mu^{ij}e_\nu^j)=0.
\label{tetrad postulate}
\end{equation}
Here, $\mathfrak{D}_\mu e_\nu^i\equiv \partial_\mu e_\nu^i - \Gamma^\rho_{~\mu\nu}e_\rho^i$ is the partial covariant derivative acting only on the two-dimensional curved indices $\mu,\nu,\dots$, but not on the Cartesian indices $i,j,\dots$.  
Eq.\eqref{tetrad postulate} is an inhomogeneous equation for $\Omega_\mu^{ij}$, thus imposing a nonzero $\Omega_\mu^{ij}$ in the laboratory frame.  As a particular solution,  we can see that the spin connection given in Eq.\eqref{2d transformation}, 
\begin{equation}
\Omega_\mu^{ij} = e_N^i\mathfrak{D}_\mu e^N_j = - e_N^j\mathfrak{D}_\mu e^N_i, 
\label{Omega formal expression for surfaces}
\end{equation}
satisfies Eq.\eqref{tetrad postulate}.  [The second equality in Eq.\eqref{Omega formal expression for surfaces} follows by (i)  $\mathfrak{D}_\mu g^{MN}=0$ since the nonzero components of $g^{MN}\equiv \frac{\partial x^M}{\partial x^i}\frac{\partial x^N}{\partial x^i}$ are $g^{\perp\perp}=1$ and $g^{\mu\nu}$, and therefore (ii)  $e_N^i\mathfrak{D}_\mu e^N_j+e_N^j\mathfrak{D}_\mu e^N_i=e_N^i\mathfrak{D}_\mu e^N_j+e^N_j\mathfrak{D}_\mu e_N^i =\partial_\mu(e_N^ie^N_j)=0$.]  
See also Ref.\cite{Kavalov1986} for 
spin connections in the Dirac theory on curved surfaces. 


From the Dirac equation $(\gamma^0\partial_t + e^\mu_i\gamma^i\nabla_\mu + m_{\rm e})\Psi=0$, with $\nabla_\mu\equiv \partial_\mu + i\Omega_\mu$, we obtain the Klein-Gordon equation,
\begin{align}
&(\gamma^0\partial_t + e^\nu_j\gamma^j\mathcal D_\nu - m_{\rm e})(\gamma^0\partial_t + e^\mu_i\gamma^i\nabla_\mu + m_{\rm e})\Psi
\notag \\
&= \left(-\partial_t^2 + \mathcal D^2 -\frac{1}{2}K - m_{\rm e}^2\right)\Psi = 0,
\label{Klein-Gordon}
\end{align}
where $\mathcal D_\lambda$ is the total covariant derivative acting both on the Cartesian and the curvilinear indices (e.g., $\mathcal D_\mu V^i_\nu\equiv \partial_\mu V^i_\nu -\Gamma^\lambda_{~\mu\nu} V^i_\lambda + \Omega_\mu^{ij}V^j_\nu$), and $\mathcal D^2\equiv g^{\mu\nu}\mathcal D_\mu \mathcal D_\nu$ is the Laplacian.  
Here, we have used that $\mathcal D_\mu$ or $\nabla_\mu$ does not act on $\gamma^i$.  (To be precise, it acts on all of the indices of $\gamma^i_{\sigma\sigma'}$ and results in zero, $\mathcal D_\mu \gamma^i= \partial_\mu\gamma^i+\Omega_\mu^{ij}\gamma^j -i\gamma^i \Omega_\mu + i\Omega_\mu \gamma^i =0$.)  
We have also used $\mathcal D_\mu e_\nu^i=0$, which is valid only for $\Omega_\mu^{ij}$ chosen as Eq.\eqref{Omega formal expression for surfaces} (if we impose also $\nabla_\mu n^i=0$).  
 Without the condition $\mathcal D_\mu e_\nu^i=0$, the derivation of the Klein-Gordon equation from the Dirac equation as in Eq.\eqref{Klein-Gordon} would be difficult.   This  is another supportive fact for the necessity of the spin connection Eq.\eqref{Omega formal expression for surfaces}.  

Let us consider the lowest-order non-relativistic limit \cite{book_Strange, book_Nishijima}.  
Separating the energy $i\partial_t$ into the rest energy and the non-relativistic energy as $i\partial_t\rightarrow  m_{\rm e} + i\partial_t$ with an approximation $i\partial_t \ll m_{\rm e}$, we have, from Eq.\eqref{Klein-Gordon}, 
\begin{equation}
i\partial_t \Psi = \left(-\frac{1}{2m_{\rm e}}\mathcal D^2 + \frac{1}{4m_{\rm e}}K\right)\Psi.
\label{nonrela sch}
\end{equation}
In the Dirac representation, where the upper two components of $\Psi$ correspond to a particle and the lower two components of $\Psi$ to an antiparticle, the $\Sigma^{ij}$ in $\Omega_\mu=\frac{1}{2}\Omega_\mu^{ij}\Sigma^{ij}$ is diagonal as $\Sigma^{ij}=\frac{1}{2}\varepsilon^{ijk}\begin{psmallmatrix}\sigma^k& 0 \\ 0 & \sigma^k\end{psmallmatrix}$.  Therefore, the upper two-components of Eq.\eqref{nonrela sch} are the usual non-relativistic Schr\"{o}dinger equation, and  it includes the nontrivial spin connection $\Omega_\mu^{ij}$ given in Eq.\eqref{Omegaij expression}.   [The term proportional to the Gaussian curvature $K$ in Eq.\eqref{nonrela sch}, which we neglect entirely in this paper as well as the ``geometric potentials'' in Refs.\cite{Jensen1971, DaCosta1981}, is akin to the electromagnetic Zeeman term $\bm\sigma\cdot \bm B$, in that the field strength $K$ of the gauge field $\Omega$ directly couples to the fermion.  $K/(4m_{\rm e})\sim 0.02~{\rm eV}$ for $K\sim 1~{\rm nm}^{-2}$ and $m_{\rm e}\sim 0.5~{\rm MeV}$.]

Next, let us consider a curve.  The discussion is completely parallel.   We employ the curvilinear coordinate system $x^M (M=s,n,b)$ as in the main text.  The Dirac action on a curve reads 
\begin{equation}
S=-i\int dtds\Psi^\dagger \gamma^0\left[\gamma^0\partial_t+e^i_s \gamma^i (\partial_s + i\Omega_s)  + m_{\rm e}\right]\Psi.
\label{Dirac curve}
\end{equation}
In order for $S$ to be real, $\Omega_s$ must satisfy
\begin{equation}
\partial_s e^i_s + \Omega_s^{ij}e_s^j=0.
\label{requirement curve}
\end{equation}
Note that $\Gamma^s_{~ss}=0$ (see Appendix \ref{app: eq7}).  Again, this is an inhomogeneous equation for $\Omega_s^{ij}$ and therefore the trivial connection $\Omega_s^{ij}=0$ is not allowed.   As a particular solution, we can see that the spin connection given in Eq.\eqref{spin connection for curves} (see Appendix \ref{app: eq7}),
\begin{equation}
\Omega_s^{ij}= e_M^i\partial_s e^M_j,
\label{Omega solution 2}
\end{equation}
satisfies the requirement Eq.\eqref{requirement curve}.  
By taking the non-relativistic limit as in the case of surfaces described above, we obtain the Schr\"{o}dinger equation
\begin{equation}
i\partial_t \Psi = -\frac{1}{2m_{\rm e}}\nabla_s \nabla_s \Psi
\end{equation}
with $\nabla_s=\partial_s + i\Omega_s$. 


Thus, to be Hermitan and to derive the Klein-Gordon equation (and the resulting Schr\"{o}dinger equation),  the Dirac theory on curves and surfaces favors the nontrivial  spin connections $\Omega$ as in Eq.\eqref{Omega formal expression for surfaces} and Eq.\eqref{Omega solution 2}, and its non-relativistic limit inherits these nontrivial spin connections.   In other words, non-relativistic theories on curved geometry should have nontrivial spin connections to be consistent with the Dirac theory.  

We can understand the discussion in this section as follows.  
In  purely non-relativistic terms such as $(\partial_\mu\psi)^2$, the spinorial index of $\psi$ can be either geometrical or non-geometrical, since the indices are contracted (summed over) only among them.  Self-consistency of such terms does not require spin connections, since, if spin connections are absent, we can always regard (or pretend) that the spinorial index of $\psi$ are non-geometrical and hence the spin connections are absent as a matter of course.   On the other hand,   in terms such as $g^{\mu\nu}e_\mu^i \gamma^i \partial_\nu\Psi$ in  the Dirac theory,  the spinorial index of $\Psi$ must be geometrical, since it is related to the spatial index ($i$ or $\mu$) via the $\gamma$-matrices.   In this case, self-consistency of the terms requires nontrivial spin connections, reflecting the geometrical nature of the spinor.

\section{Details of Eq.\eqref{Omegaij expression}}\label{app: eq3}

Here, we calculate some of the components of $\Gamma^M_{~NL}$ and then derive Eq.\eqref{Omegaij expression}.  
First, the Cartesian coordinates $x^i$ and the curvilinear coordinates $x^M=(x^\mu, x^\perp)$ are related by, as in Ref.\cite{DaCosta1981},
\begin{equation}
x^i(x^M) = r^i(x^\mu) + x^\perp n^i(x^\mu),
\end{equation}
where $r^i(x^\mu)$ specifies a point on a surface, and $n^i(x^\mu)$ is the surface normal at each point $r^i(x^\mu)$.   
Hereafter, any quantity is eventually evaluated just on the surface, i.e., at $x^\perp=0$.  

It is very convenient to give a name to $\partial_i n^j$.  We call it the extrinsic curvature tensor or the second fundamental form\cite{book_Carroll, book_Poisson, book_David, book_doCarmo} $K_i^{~j}$:
\begin{equation}
K_i^{~j} \equiv \partial_i n^j,
\end{equation}
where we have extended the domain of $n^j$ infinitesimally off the surface in a way such that $\partial_\perp n^j=0$.  
We can easily see that $n^iK_i^{~j}=\partial_\perp n^j=0$  and $K_i^{~j}n^j=0$.  Therefore, when we express $K$ by the curvilinear indices as $K_M^{~N}=\nabla^{({\rm flat})}_M n^N$,  then $n^M K_M^{~N}=K_\perp^{~N}=0$ and $K_M^{~N}n_N=K_M^{~\perp}=0$  (note that $n^M=\delta^{M}_{\perp}$ and $n_M=\delta_{M}^{\perp}$).  Therefore, 
\begin{equation}
\mbox{only the tangential components $K_\mu^{~\nu}$ are nonzero.}
\label{nonzero}
\end{equation}
Moreover, since $\nabla^{({\rm flat})}_M n^N=\partial_Mn^N+\Gamma^N_{~ML}n^L=\Gamma^N_{~M\perp}$ and $\nabla^{({\rm flat})}_M n_N=\partial_M n_N - \Gamma^L_{~MN}n_L= - \Gamma^\perp_{~MN}$, we have 
\begin{equation}
\Gamma^N_{~M\perp} = K_M^{~N},~~~\Gamma^\perp_{~MN}= - K_{MN}.
\end{equation}
From this and Eq.\eqref{nonzero}, we see that the components of $\Gamma^M_{~NL}$ with more than one normal index ($\perp$), such as $\Gamma^\perp_{~N\perp}$, are zero.  The nonzero components with the normal index are 
\begin{equation}
\Gamma^\nu_{~\mu\perp} = K_\mu^{~\nu},~~~\Gamma^\perp_{~\mu\nu}=-K_{\mu\nu}.
\label{Gamma and K}
\end{equation}
Let us next consider the contraction of $K$ with $e^i_M\equiv \partial x^i/\partial x^M$ and $e^M_i \equiv \partial x^M/\partial x^i$.  This is easy, because $K$ is a tensor and multiplication of it with $e^i_M$ or $e^M_i$ just changes its corresponding index.  For example, 
\begin{align}
K_{\mu N}e^{N}_j=K_{\mu\nu}e^{\nu}_j = K_\mu^{~j} = \nabla_\mu^{({\rm flat})} n^j = \partial_\mu n^j. 
\label{K and e}
\end{align}
(Note that we do not have to make a distinction between raised and lowered Cartesian indices.)  From Eq.\eqref{Gamma and K},  Eq.\eqref{K and e}, and $e^i_\perp=e^\perp_i=n^i$, the detail of Eq.\eqref{Omegaij expression} reads 
\begin{align}
\Omega_\mu^{ij} =& - e^i_\perp \Gamma^\perp_{~\mu\lambda}e^{\lambda}_j - e^i_\nu \Gamma^\nu_{~\mu\perp}e^{\perp}_j
\notag \\
=& n^i K_{\mu\lambda}e^{\lambda}_j - n^j K_\mu^{~\nu}e^i_\nu 
\notag \\
=& n^i\partial_\mu n^j - n^j\partial_\mu n^i.
\end{align}
We can derive this expression more directly from Eq.\eqref{2d transformation} or Eq.\eqref{Omega formal expression for surfaces} as follows.  A useful identity is 
\begin{equation}
\mathfrak{D}_\mu e_\nu^i\equiv \partial_\mu e_\nu^i-\Gamma^\rho_{~\mu\nu}e_\rho^i=-K_{\mu\nu}n^i,
\end{equation}
which follows from (i) $e_\rho^i\mathfrak{D}_\mu e_\nu^i=0$ by using $\Gamma^{\lambda}_{~\mu\nu}=e^{\lambda}_k\partial_\mu e_\nu^k$, and (ii) $K_{\mu\nu}=-n^i\partial_\mu e_\nu^i=-n^i\mathfrak{D}_\mu e_\nu^i$ by applying $\partial_\mu$ on $n^ie_\nu^i=0$.  Then, 
\begin{align}
\Omega_\mu^{ij} =& g^{\nu\rho}e_\nu^i \mathfrak{D}_\mu e_\rho^j + n^i\partial_\mu n^j
\notag \\
=& -e_\nu^i K_{\mu}^{~\nu}n^j + n^i\partial_\mu n^j
\notag \\
=& -\partial_\mu n^i n^j + n^i\partial_\mu n^j.
\end{align}

\section{Details of Eq.\eqref{intrinsic spin connection}}\label{app: eq5}

A useful identity to show Eq.\eqref{intrinsic spin connection} is 
\begin{equation}
iU\partial_\mu U^\dagger = -(\bm n\times \partial_\mu \bm n)\cdot \frac{\bm \sigma}{2} + (1-\cos\theta)\partial_\mu\phi \left( \bm n\cdot \frac{\bm \sigma}{2}\right).
\label{identity}
\end{equation}
Here, $U$ is defined as a SU(2) matrix satisfying $U^\dagger \bm n\cdot \bm \sigma U=\sigma^3$, whose particular form we choose is 
\begin{equation}
U=i\bm m\cdot \bm \sigma ~~~\mbox{where}~~~\bm m=
\begin{pmatrix}
\sin\frac{\theta}{2}\cos\phi \\
\sin\frac{\theta}{2}\sin\phi \\
\cos\frac{\theta}{2}
\end{pmatrix}
\label{U}
\end{equation}  
with $\bm n=(\sin\theta\cos\phi, \sin\theta\sin\phi, \cos\theta)$.  The definition of $U$ has ambiguity up to the U(1) rotation around the $\sigma^3$-axis, $U\rightarrow U\exp(i\chi\sigma^3/2)$ with $\chi$ an arbitrary function.  This U(1) indefiniteness corresponds to the $U(1)$ gauge transformation on the `monopole gauge field' $(1-\cos\theta)\partial_\mu\phi$ in the second term in Eq.\eqref{identity}.  When we parameterize a general SU(2) matrix $\mathcal U$  by the Euler angles $(\theta,\phi,\psi)$,
\begin{equation}
\mathcal U \equiv e^{-i\phi\frac{\sigma^3}{2}}e^{-i\theta\frac{\sigma^2}{2}}e^{-i\psi\frac{\sigma^3}{2}},
\end{equation}
the matrix $U$ in Eq.\eqref{U} corresponds to the restriction $\psi=-\phi-\pi$, that is, $\mathcal U|_{\psi=-\phi-\pi}=U$.

We define a rotation matrix $R_{ia}$ by $R_{ia}\sigma^a=U^\dagger \sigma^i U$, whose explicit form  is $R_{ia}=2m^im^a-\delta^{ia}$ for $U$ chosen as Eq.\eqref{U}.  Then, the detail of Eq.\eqref{intrinsic spin connection} is   
\begin{align}
\omega_\mu =& U^\dagger \Omega_\mu U - iU^\dagger \partial_\mu U
\notag \\
=&R_{ia}\Big[(\bm n\times \partial_\mu \bm n)^i \frac{\sigma^a}{2}-(\bm n\times \partial_\mu \bm n)^i\frac{\sigma^a}{2} 
\notag \\
&~~~~~~~+ (1-\cos\theta)\partial_\mu\phi~ n^i \frac{\sigma^a}{2}\Big]
\notag \\
=& (1-\cos\theta)\partial_\mu \phi ~\frac{\sigma^3}{2},
\label{transition}
\end{align}
where we have used $R_{ia}n^i=\delta^{a3}$.

The calculation above can be rephrased for the case of vectors.  When we define  $\omega_\mu^a$ as 
\begin{equation}
\partial_\mu V^i - \varepsilon_{ijk}\Omega_\mu^j V^k 
= R_{ia}\left(\partial_\mu V^a - \varepsilon_{abc}\omega_\mu^b V^c\right) 
\end{equation}
with $V^i=R_{ia}V^a$, then 
\begin{align}
\omega_\mu^a =& R_{ia}\Omega_\mu^i + \frac{1}{2}\varepsilon_{abc}(R^{-1}\partial_\mu R)_{bc}
\notag \\
=& \delta^{a3}(1-\cos\theta)\partial_\mu\phi.
\end{align}

\section{Details of spin-dependent magnetic field}\label{app: pseudomagnetic field}

Here, we show that the ``pseudomagnetic field'' $\pm\bm B_{\rm m}$ calculated from the spin connection in Eq.\eqref{intrinsic spin connection} is given as $\bm B_{\rm m}=(K/2)\bm n$. 

We can roughly expect $\bm B_{\rm m}=(K/2)\bm n$ as follows.  The pseudomagnetic field is a quantity of the second-order derivative.  In order to evaluate it, we may approximate the portion of the surface near a point in interest by a sphere, whose radius is $1/\sqrt{K}$ with $K$ the (Gaussian) curvature of the point.  Within this approximation, the normal vectors $\bm n$ of the surface are the radial unit vectors of the sphere, and the intrinsic spin connection $\omega_\mu =(1-\cos\theta)\partial_\mu\phi\frac{\sigma^3}{2}$ can be regarded as the electromagnetic gauge fields of magnetic monopoles with charges $\pm \frac{1}{2}$ put at the center of the sphere.  Generally, the magnetic field at a point $\bm r$, yielded by a monopole with charge $q$ put at the origin, is given by $\frac{q}{|\bm r|^2}\frac{\bm r}{|\bm r|}$.  Therefore, the pseudomagnetic fields are given by $\pm\bm B_{\rm m}=\pm(K/2)\bm n$.    

We calculate the pseudomagnetic fields more rigorously below. 
First, the two-dimensional Riemann tensor $R^{\mu}_{~\nu\rho\sigma}$ on a surface is defined by
\begin{equation}
R^{\mu}_{~\nu\rho\sigma} \equiv \partial_\rho \Gamma^\mu_{~\nu\sigma} + \Gamma^\mu_{~\rho\lambda}\Gamma^\lambda_{~\nu\sigma} - (\rho\leftrightarrow \sigma).
\label{Riemann 1}
\end{equation}
Comparing this with an obvious identity 
\begin{equation}
0= \partial_\rho \Gamma^\mu_{~\nu\sigma} + \Gamma^\mu_{~\rho L}\Gamma^L_{~\nu\sigma} - (\rho\leftrightarrow \sigma)
\end{equation}
(giving  the tangential components of the Riemann tensor in the flat three-dimensional space, which are of course zero), we have
\begin{align}
R^{\mu}_{~\nu\rho\sigma} =&  - \Gamma^\mu_{~\rho \perp}\Gamma^\perp_{~\nu\sigma}
- (\rho\leftrightarrow \sigma)
\notag \\
=& K^\mu_{~\rho} K_{\nu\sigma} - K^\mu_{~\sigma} K_{\nu\rho},  
\end{align}
where $K^\mu_{~\nu}$ is the extrinsic curvature tensor (see Appendix \ref{app: eq3}).  Then, the Ricci scalar $R$ is given by  
\begin{align}
R\equiv & R^{\mu\nu}_{~~~\mu\nu}
\notag \\
=& (K^\mu_{~\mu})^2 - K^\mu_{~\nu}K^\nu_{~\mu}
\notag \\
=& 2K,
\label{Ricci scalar}
\end{align}
where $K\equiv {\rm det}(K^\mu_{~\nu})$ is the Gaussian curvature, and we have used an identity ${\rm det}A=\frac{1}{2}[({\rm tr}A)^2-{\rm tr}(A^2)]$ valid for any $2\times 2$ matrix $A$.  

The Riemann tensor can be defined in  another way as 
\begin{align}
R_{\mu\nu}\psi \equiv& -i[\mathcal D_\mu, \mathcal D_\nu]\psi
\notag \\
=& \left(\partial_\mu \Omega_\nu - \partial_\nu \Omega_\mu + i[\Omega_\mu, \Omega_\nu]\right)\psi
\notag \\
=& [\bm n\cdot (\partial_\mu \bm n\times \partial_\nu \bm n)] \left(\bm n \cdot \frac{\bm \sigma}{2}\right)\psi,
\label{Riemann 2}
\end{align}
(we do not use $R_{\mu\nu}$ to describe the Ricci tensor) where $\mathcal D_\mu$ is the total covariant derivative acting both on the curvilinear and the Cartesian indices (e.g., $\mathcal D_\mu V^i_\nu = \partial_\mu V_\nu^i - \Gamma^\lambda_{~\mu\nu}V_\lambda^i + \Omega_\mu^{ij}V_\nu^j$), and $\Omega_\mu=\bm \Omega_\mu\cdot \frac{\bm \sigma}{2}$ is the spin connection Eq.\eqref{covariant derivative} in the laboratory frame.  
We have used $\partial_\mu \bm n \times \partial_\nu \bm n = [\bm n\cdot (\partial_\mu \bm n \times \partial_\nu \bm n)]\bm n$.  
Defining $R_{\mu\nu}^i\frac{\sigma^i}{2}\equiv R_{\mu\nu}$ and $R_{\mu\nu}^{ij} \equiv \varepsilon^{ijk}R_{\mu\nu}^k$, the two ways of the definition of the Riemann tensor, Eq.\eqref{Riemann 1} and Eq.\eqref{Riemann 2}, are related by
\begin{equation}
R_{\mu\nu\rho\sigma}=e_\rho^ie_\sigma^jR_{\mu\nu}^{ij},~~ R_{\mu\nu}^{ij} = e_{\rho}^ie_{\sigma}^j R_{\mu\nu}^{~~~\rho\sigma}, 
\label{relation}
\end{equation}
where $e_\mu^i\equiv \partial x^i/\partial x^\mu$ and we have implicitly used $n^iR_{\mu\nu}^{ij}=0$.  The relation Eq.\eqref{relation} follows from  $[\mathcal D_\mu, \mathcal D_\nu]e_\rho^i = R_{\mu\nu}^{ij}e_\rho^j - R^\lambda_{~\rho\mu\nu}e_\lambda^i$, which vanishes due to  $\mathcal D_\mu e_\nu^i=0$.    Then, from Eq.\eqref{Riemann 2} and Eq.\eqref{relation}, we have 
\begin{align}
R =& e^\mu_i e^\nu_j R_{\mu\nu}^{ij}
\notag \\
=& [\bm n\cdot (\bm e^\mu\times \bm e^\nu)] [\bm n\cdot (\partial_\mu \bm n \times \partial_\nu \bm n)]
\notag \\
=&\frac{2}{\sqrt{g}}\bm n\cdot (\partial_1 \bm n \times \partial_2 \bm n),
\label{R and solid angle}
\end{align}
where we have used $\bm e^1\times \bm e^2=\frac{1}{\sqrt{g}}\bm n$ (or $\bm e_1\times \bm e_2=\sqrt{g}\bm n$).  Eq.\eqref{R and solid angle} means, from Eq.\eqref{Ricci scalar}, that the Gaussian curvature $K\delta x^1 \delta x^2$ with 
\begin{equation}
K=\frac{1}{\sqrt{g}}\bm n\cdot (\partial_1 \bm n \times \partial_2 \bm n)
\label{solid angle}
\end{equation}
at a point $x=(x^1, x^2)$ is given by the solid angle spanned by the normal vectors at $x$, $(x^1+\delta x^1, x^2)$ and $(x^1, x^2+\delta x^2)$. 

The Riemann tensor $\widetilde R_{\mu\nu}$ in the intrinsic frame is given by 
\begin{align}
\widetilde R_{\mu\nu} =& \partial_\mu \omega_\nu - \partial_\nu \omega_\mu
\notag \\
=&\bm n\cdot (\partial_\mu \bm n \times \partial_\nu \bm n)\frac{\sigma^3}{2}
\label{Riemann 3}
\end{align}
with $\omega_\mu\equiv (1-\cos\theta)\partial_\mu\phi(\sigma^3/2)$ as in Eq.\eqref{intrinsic spin connection}.   Then, the pseudomagnetic field $B_{\rm m}^i$ is defined by $B^i_{\rm m}\equiv \frac{1}{2}\varepsilon^{ijk}F_{jk}$ with 
 $F_{ij}\equiv e^\mu_i e^\nu_j F_{\mu\nu}$ and $F_{\mu\nu}\sigma^3 \equiv \widetilde R_{\mu\nu}$.  
Then, from Eq.\eqref{solid angle} and Eq.\eqref{Riemann 3}, we have
\begin{align}
\bm B_{\rm m} =& \frac{1}{4}[\bm n\cdot (\partial_\mu \bm n \times \partial_\nu \bm n)](\bm e^\mu \times \bm e^\nu)
\notag \\
=& \frac{K}{2}\bm n.
\end{align}

\section{Details of Eq.\eqref{spin connection for curves}} \label{app: eq7}

Let $\bm \gamma(s)$ be the position of a point on a curve  parametrized by its arclength $s$.   The curve determines by itself a natural orthonormal frame  consisting of unit vectors $(\bm T, \bm N, \bm B)$, where $\bm T(s)\equiv \bm \gamma'(s)$, $\bm N(s)\equiv \bm T'(s)/|\bm T' (s)|$ and $\bm B(s) \equiv \bm T(s)\times \bm N(s)$.  With these orthonormal vectors, we define a curvilinear coordinate system $x^M$ with $M=s,n,b$ by 
\begin{equation}
x^i(x^M)=\gamma^i(x^s)+ x^n N^i(x^s) + x^b B^i(x^s).   
\end{equation}
Hereafter, all quantities are eventually evaluated just on the curve, i.e., at $x^n=x^b=0$.  

The nonzero components of the geometric connection $\Gamma^M_{~NL}$ in this curvilinear coordinate system are 
\begin{equation}
-\Gamma^s_{~sn}=\Gamma^n_{~ss}=\kappa,~~-\Gamma^n_{~sb}=\Gamma^b_{~sn}=\tau, 
\label{nonzero components for curves}
\end{equation} 
where $\kappa\equiv \bm N\cdot \bm T'$ and $\tau \equiv -\bm N\cdot \bm B'$ are the curvature and the torsion of the curve, respectively.  These components can be calculated directly by an expression 
\begin{equation}
\Gamma^M_{~NL} = \frac{1}{2}G^{MP}\left(
  \frac{\partial G_{PN}}{\partial x^L}
+\frac{\partial G_{PL}}{\partial x^N}
-\frac{\partial G_{NL}}{\partial x^P}
\right),
\end{equation}
where $G_{MN}\equiv (\partial x^i/\partial x^M)(\partial x^i/\partial x^N)$ is the metric defined in the three-dimensional space.  We  can also calculate $\Gamma^M_{~NL}$ more quickly by deriving the Euler-Lagrange equation for $L=\frac{1}{2}G_{MN}(X)\dot X^M \dot X^N$ for the position $X^M(t)$ of a point particle at time $t$, as explained in Chap. 2 in Ref.\cite{book_FosterNightingale}.  

Defining $e^i_M\equiv \partial x^i/\partial x^M$ and $e^M_i\equiv \partial x^M/\partial x^i$, we can easily see that $e^i_s=e^s_i=T^i$, $e^i_n=e^n_i=N^i$ and $e^i_b=e^b_i=B^i$.  With all of this setup, a procedure similar to the case of surfaces gives an expression of the spin connection $\Omega_s^{ij}$ on curves.  
Let us start from an obvious identity
\begin{equation}
0 = e^{i}_{N} \Gamma^N_{~s L} e^{L}_j 
+ e^{i}_N\partial_s e^N_j,
\end{equation}
which simply states that the spin connection for the flat derivative is zero, $\nabla_\mu^{({\rm flat})}V^i=\partial_\mu V^i$.  Then, the spin connection for the curved derivative is given by truncating all of the components of $\Gamma^M_{~NL}$ other than $\Gamma^s_{~ss}$,
\begin{equation}
\Omega_s^{ij} =  e^{i}_{N}\partial_s e^N_j 
\label{Omega formal expression for curves}
\end{equation}
(note that $\Gamma^s_{~ss}=0$).  
By comparing these equations,  we have, by Eq.\eqref{nonzero components for curves},
\begin{align}
\Omega_s^{ij}=& -\left(e^i_s\Gamma^s_{~sn}e^n_j + e^i_n\Gamma^n_{~ss}e^s_j + e^i_n\Gamma^n_{~sb}e^b_j + e^i_b\Gamma^b_{~sn}e^n_j
\right)
\notag \\
=& \kappa (T^iN^j-N^iT^j) + \tau(N^iB^j-B^iN^j), 
\end{align}
which gives Eq.\eqref{spin connection for curves} by $\Omega_s^i \equiv \frac{1}{2}\varepsilon_{ijk}\Omega_s^{jk}$.  

We can derive this expression more directly from Eq.\eqref{Omega formal expression for curves}, 
\begin{align}
\Omega_s^{ij} =& T^i T'^j + N^i N'^j + B^i B'^j
\notag \\
=& \kappa (T^iN^j-N^iT^j) + \tau(N^iB^j-B^iN^j),
\end{align}
where we have used the Frenet--Serret formula.

\section{Spin connections on curves vanish in the intrinsic frame.}\label{app: gauge trf for curves}

As described in the main text, the spin connection for a curve  is $\bm \Omega_s = \kappa \bm B + \tau \bm T$ in the laboratory frame.  Here, we show that this spin connection vanishes when we go to the intrinsic frame.  

The relevant SO(3) transformation matrix is 
\begin{equation}
\mathcal R_{ia}\equiv (\bm N, \bm B, \bm T)_{ia}.
\end{equation}
It is obvious that
\begin{equation}
\mathcal R_{ia}N^i=\delta^{a1},~~ \mathcal R_{ia}B^i=\delta^{a2},~~\mathcal R_{ia}T^i=\delta^{a3},  
\end{equation}
which means that $\mathcal R_{ia}$ is the rotation matrix from the laboratory (Cartesian) frame to the intrinsic frame, in the latter of which $\bm N$, $\bm B$ and $\bm T$ are the basis vectors.  

The spin connection $\bm \Omega_s$ on a curve is simply the `angular velocity' of $\mathcal R$:
\begin{equation}
\Omega_s^i = \frac{1}{2}\varepsilon^{ijk}\left(\mathcal R\frac{d}{ds}\mathcal R^{-1}\right)_{jk},
\label{Omegas as angular velocity}
\end{equation}
which is equivalent to $d\mathcal R_{ia}/ds=\varepsilon^{ijk}\Omega_s^j\mathcal R_{ka}$, or the Frenet--Serret formulas $d \bm X/ds = \bm \Omega_s\times \bm X$ for $\bm X=\bm T, \bm N, \bm B$.   
Eq.\eqref{Omegas as angular velocity} means that $\mathcal R_{ia}$ transforms $\Omega_s^i$ to zero, as 
\begin{equation}
\mathcal R_{ia}\Omega_s^i + \frac{1}{2}\varepsilon^{abc}\left(\mathcal R^{-1}\frac{d}{ds}\mathcal R\right)_{bc} = 0.
\end{equation}

These can be rephrased by using the SU(2) matrix $\mathcal U$ defined as $\mathcal U^\dagger \sigma^i \mathcal U=\mathcal R_{ia}\sigma^a$.  
From an identity 
\begin{equation}
-i\mathcal U \frac{d}{ds}\mathcal U^\dagger 
=  \frac{1}{2}\varepsilon^{ijk}\left(\mathcal R\frac{d}{ds}\mathcal R^{-1}\right)_{jk}\frac{\sigma^i}{2},
\end{equation}
we have 
\begin{equation}
\bm \Omega_s\cdot \frac{\bm \sigma}{2} = -i\mathcal U \frac{d}{ds}\mathcal U^\dagger,
\end{equation}
which is transformed to zero in the intrinsic frame as
\begin{equation}
\mathcal U^\dagger \left(\bm \Omega_s\cdot \frac{\bm \sigma}{2}\right)\mathcal U -i\mathcal U^\dagger \frac{d}{ds}\mathcal U=0. 
\end{equation}


\bibliographystyle{apsrev4-1}
\bibliography{spin_connection}

\begin{thebibliography}{95}%
\makeatletter
\providecommand \@ifxundefined [1]{%
 \@ifx{#1\undefined}
}%
\providecommand \@ifnum [1]{%
 \ifnum #1\expandafter \@firstoftwo
 \else \expandafter \@secondoftwo
 \fi
}%
\providecommand \@ifx [1]{%
 \ifx #1\expandafter \@firstoftwo
 \else \expandafter \@secondoftwo
 \fi
}%
\providecommand \natexlab [1]{#1}%
\providecommand \enquote  [1]{``#1''}%
\providecommand \bibnamefont  [1]{#1}%
\providecommand \bibfnamefont [1]{#1}%
\providecommand \citenamefont [1]{#1}%
\providecommand \href@noop [0]{\@secondoftwo}%
\providecommand \href [0]{\begingroup \@sanitize@url \@href}%
\providecommand \@href[1]{\@@startlink{#1}\@@href}%
\providecommand \@@href[1]{\endgroup#1\@@endlink}%
\providecommand \@sanitize@url [0]{\catcode `\\12\catcode `\$12\catcode
  `\&12\catcode `\#12\catcode `\^12\catcode `\_12\catcode `\%12\relax}%
\providecommand \@@startlink[1]{}%
\providecommand \@@endlink[0]{}%
\providecommand \url  [0]{\begingroup\@sanitize@url \@url }%
\providecommand \@url [1]{\endgroup\@href {#1}{\urlprefix }}%
\providecommand \urlprefix  [0]{URL }%
\providecommand \Eprint [0]{\href }%
\providecommand \doibase [0]{http://dx.doi.org/}%
\providecommand \selectlanguage [0]{\@gobble}%
\providecommand \bibinfo  [0]{\@secondoftwo}%
\providecommand \bibfield  [0]{\@secondoftwo}%
\providecommand \translation [1]{[#1]}%
\providecommand \BibitemOpen [0]{}%
\providecommand \bibitemStop [0]{}%
\providecommand \bibitemNoStop [0]{.\EOS\space}%
\providecommand \EOS [0]{\spacefactor3000\relax}%
\providecommand \BibitemShut  [1]{\csname bibitem#1\endcsname}%
\let\auto@bib@innerbib\@empty
\bibitem [{\citenamefont {Sakurai}(1994)}]{book_Sakurai}%
  \BibitemOpen
  \bibfield  {author} {\bibinfo {author} {\bibfnamefont {J.~J.}\ \bibnamefont
  {Sakurai}},\ }\href@noop {} {\emph {\bibinfo {title} {{Modern Quantum
  Mechanics}}}}\ (\bibinfo  {publisher} {Addison-Wesley},\ \bibinfo {year}
  {1994})\BibitemShut {NoStop}%
\bibitem [{\citenamefont {Carroll}(2003)}]{book_Carroll}%
  \BibitemOpen
  \bibfield  {author} {\bibinfo {author} {\bibfnamefont {S.}~\bibnamefont
  {Carroll}},\ }\href@noop {} {\emph {\bibinfo {title} {{Spacetime and
  Geometry}}}}\ (\bibinfo  {publisher} {Addison-Wesley},\ \bibinfo {year}
  {2003})\BibitemShut {NoStop}%
\bibitem [{\citenamefont {Nakahara}(2003)}]{book_Nakahara}%
  \BibitemOpen
  \bibfield  {author} {\bibinfo {author} {\bibfnamefont {M.}~\bibnamefont
  {Nakahara}},\ }\href@noop {} {\emph {\bibinfo {title} {{Geometry, Topology
  and Physics}}}},\ \bibinfo {edition} {2nd}\ ed.\ (\bibinfo  {publisher} {CRC
  Press},\ \bibinfo {year} {2003})\BibitemShut {NoStop}%
\bibitem [{\citenamefont {Kamien}(2002)}]{Kamien2002}%
  \BibitemOpen
  \bibfield  {author} {\bibinfo {author} {\bibfnamefont {R.~D.}\ \bibnamefont
  {Kamien}},\ }\href {\doibase 10.1103/RevModPhys.74.953} {\bibfield  {journal}
  {\bibinfo  {journal} {Rev. Mod. Phys.}\ }\textbf {\bibinfo {volume} {74}},\
  \bibinfo {pages} {953} (\bibinfo {year} {2002})}\BibitemShut {NoStop}%
\bibitem [{\citenamefont {David}(2004)}]{book_David}%
  \BibitemOpen
  \bibfield  {author} {\bibinfo {author} {\bibfnamefont {F.}~\bibnamefont
  {David}},\ }\href@noop {} {\emph {\bibinfo {title} {{Statistical Mechanics of
  Membranes and Surfaces}}}},\ \bibinfo {edition} {2nd}\ ed.,\ edited by\
  \bibinfo {editor} {\bibfnamefont {D.}~\bibnamefont {Nelson}}, \bibinfo
  {editor} {\bibfnamefont {T.}~\bibnamefont {Piran}}, \ and\ \bibinfo {editor}
  {\bibfnamefont {S.}~\bibnamefont {Weinberg}}\ (\bibinfo  {publisher} {World
  Scientific},\ \bibinfo {year} {2004})\BibitemShut {NoStop}%
\bibitem [{\citenamefont {Green}\ \emph {et~al.}(1987)\citenamefont {Green},
  \citenamefont {Schwarz},\ and\ \citenamefont {Witten}}]{book_GSW}%
  \BibitemOpen
  \bibfield  {author} {\bibinfo {author} {\bibfnamefont {M.~B.}\ \bibnamefont
  {Green}}, \bibinfo {author} {\bibfnamefont {J.~H.}\ \bibnamefont {Schwarz}},
  \ and\ \bibinfo {author} {\bibfnamefont {E.}~\bibnamefont {Witten}},\
  }\href@noop {} {\emph {\bibinfo {title} {{Superstring Theory}}}}\ (\bibinfo
  {publisher} {Cambridge University Press},\ \bibinfo {year}
  {1987})\BibitemShut {NoStop}%
\bibitem [{\citenamefont {Parker}\ and\ \citenamefont
  {Toms}(2009)}]{book_ParkerToms}%
  \BibitemOpen
  \bibfield  {author} {\bibinfo {author} {\bibfnamefont {L.}~\bibnamefont
  {Parker}}\ and\ \bibinfo {author} {\bibfnamefont {D.}~\bibnamefont {Toms}},\
  }\href@noop {} {\emph {\bibinfo {title} {{Quantum Field Theory in Curved
  Spacetime: Quantized Fields and Gravity}}}}\ (\bibinfo  {publisher}
  {Cambridge University Press},\ \bibinfo {year} {2009})\BibitemShut {NoStop}%
\bibitem [{\citenamefont {Barnett}(1915)}]{Barnett1915}%
  \BibitemOpen
  \bibfield  {author} {\bibinfo {author} {\bibfnamefont {S.~J.}\ \bibnamefont
  {Barnett}},\ }\href {\doibase 10.1103/PhysRev.6.239} {\bibfield  {journal}
  {\bibinfo  {journal} {Phys. Rev.}\ }\textbf {\bibinfo {volume} {6}},\
  \bibinfo {pages} {239} (\bibinfo {year} {1915})}\BibitemShut {NoStop}%
\bibitem [{\citenamefont {Barnett}(1935)}]{Barnett1935}%
  \BibitemOpen
  \bibfield  {author} {\bibinfo {author} {\bibfnamefont {S.~J.}\ \bibnamefont
  {Barnett}},\ }\href {\doibase 10.1103/RevModPhys.7.129} {\bibfield  {journal}
  {\bibinfo  {journal} {Rev. Mod. Phys.}\ }\textbf {\bibinfo {volume} {7}},\
  \bibinfo {pages} {129} (\bibinfo {year} {1935})}\BibitemShut {NoStop}%
\bibitem [{\citenamefont {Stockhofe}\ and\ \citenamefont
  {Schmelcher}(2014)}]{Stockhofe2014}%
  \BibitemOpen
  \bibfield  {author} {\bibinfo {author} {\bibfnamefont {J.}~\bibnamefont
  {Stockhofe}}\ and\ \bibinfo {author} {\bibfnamefont {P.}~\bibnamefont
  {Schmelcher}},\ }\href {\doibase 10.1103/PhysRevA.89.033630} {\bibfield
  {journal} {\bibinfo  {journal} {Phys. Rev. A}\ }\textbf {\bibinfo {volume}
  {89}},\ \bibinfo {pages} {033630} (\bibinfo {year} {2014})}\BibitemShut
  {NoStop}%
\bibitem [{\citenamefont {Meijer}\ \emph {et~al.}(2002)\citenamefont {Meijer},
  \citenamefont {Morpurgo},\ and\ \citenamefont {Klapwijk}}]{Meijer2002}%
  \BibitemOpen
  \bibfield  {author} {\bibinfo {author} {\bibfnamefont {F.~E.}\ \bibnamefont
  {Meijer}}, \bibinfo {author} {\bibfnamefont {A.~F.}\ \bibnamefont
  {Morpurgo}}, \ and\ \bibinfo {author} {\bibfnamefont {T.~M.}\ \bibnamefont
  {Klapwijk}},\ }\href {\doibase 10.1103/PhysRevB.66.033107} {\bibfield
  {journal} {\bibinfo  {journal} {Phys. Rev. B}\ }\textbf {\bibinfo {volume}
  {66}},\ \bibinfo {pages} {033107} (\bibinfo {year} {2002})}\BibitemShut
  {NoStop}%
\bibitem [{\citenamefont {Wen}\ and\ \citenamefont {Zee}(1992)}]{Wen1992}%
  \BibitemOpen
  \bibfield  {author} {\bibinfo {author} {\bibfnamefont {X.~G.}\ \bibnamefont
  {Wen}}\ and\ \bibinfo {author} {\bibfnamefont {A.}~\bibnamefont {Zee}},\
  }\href {\doibase 10.1103/PhysRevLett.69.953} {\bibfield  {journal} {\bibinfo
  {journal} {Phys. Rev. Lett.}\ }\textbf {\bibinfo {volume} {69}},\ \bibinfo
  {pages} {953} (\bibinfo {year} {1992})}\BibitemShut {NoStop}%
\bibitem [{\citenamefont {Fr{\"{o}}hlich}\ and\ \citenamefont
  {Studer}(1993)}]{Froehlich1993}%
  \BibitemOpen
  \bibfield  {author} {\bibinfo {author} {\bibfnamefont {J.}~\bibnamefont
  {Fr{\"{o}}hlich}}\ and\ \bibinfo {author} {\bibfnamefont {U.~M.}\
  \bibnamefont {Studer}},\ }\href {\doibase 10.1103/RevModPhys.65.733}
  {\bibfield  {journal} {\bibinfo  {journal} {Rev. Mod. Phys.}\ }\textbf
  {\bibinfo {volume} {65}},\ \bibinfo {pages} {733} (\bibinfo {year}
  {1993})}\BibitemShut {NoStop}%
\bibitem [{\citenamefont {Jensen}\ and\ \citenamefont
  {Koppe}(1971)}]{Jensen1971}%
  \BibitemOpen
  \bibfield  {author} {\bibinfo {author} {\bibfnamefont {H.}~\bibnamefont
  {Jensen}}\ and\ \bibinfo {author} {\bibfnamefont {H.}~\bibnamefont {Koppe}},\
  }\href {\doibase 10.1016/0003-4916(71)90031-5} {\bibfield  {journal}
  {\bibinfo  {journal} {Ann. Phys.}\ }\textbf {\bibinfo {volume} {63}},\
  \bibinfo {pages} {586} (\bibinfo {year} {1971})}\BibitemShut {NoStop}%
\bibitem [{\citenamefont {da~Costa}(1981)}]{DaCosta1981}%
  \BibitemOpen
  \bibfield  {author} {\bibinfo {author} {\bibfnamefont {R.~C.~T.}\
  \bibnamefont {da~Costa}},\ }\href {\doibase 10.1103/PhysRevA.23.1982}
  {\bibfield  {journal} {\bibinfo  {journal} {Phys. Rev. A}\ }\textbf {\bibinfo
  {volume} {23}},\ \bibinfo {pages} {1982} (\bibinfo {year}
  {1981})}\BibitemShut {NoStop}%
\bibitem [{\citenamefont {Gonz{\'{a}}lez}\ \emph {et~al.}(1992)\citenamefont
  {Gonz{\'{a}}lez}, \citenamefont {Guinea},\ and\ \citenamefont
  {Vozmediano}}]{Gonzalez1992}%
  \BibitemOpen
  \bibfield  {author} {\bibinfo {author} {\bibfnamefont {J.}~\bibnamefont
  {Gonz{\'{a}}lez}}, \bibinfo {author} {\bibfnamefont {F.}~\bibnamefont
  {Guinea}}, \ and\ \bibinfo {author} {\bibfnamefont {M.~A.~H.}\ \bibnamefont
  {Vozmediano}},\ }\href {\doibase 10.1103/PhysRevLett.69.172} {\bibfield
  {journal} {\bibinfo  {journal} {Phys. Rev. Lett.}\ }\textbf {\bibinfo
  {volume} {69}},\ \bibinfo {pages} {172} (\bibinfo {year} {1992})}\BibitemShut
  {NoStop}%
\bibitem [{\citenamefont {Kane}\ and\ \citenamefont {Mele}(1997)}]{Kane1997}%
  \BibitemOpen
  \bibfield  {author} {\bibinfo {author} {\bibfnamefont {C.~L.}\ \bibnamefont
  {Kane}}\ and\ \bibinfo {author} {\bibfnamefont {E.~J.}\ \bibnamefont
  {Mele}},\ }\href {\doibase 10.1103/PhysRevLett.78.1932} {\bibfield  {journal}
  {\bibinfo  {journal} {Phys. Rev. Lett.}\ }\textbf {\bibinfo {volume} {78}},\
  \bibinfo {pages} {1932} (\bibinfo {year} {1997})}\BibitemShut {NoStop}%
\bibitem [{\citenamefont {Lee}(2009)}]{Lee2009}%
  \BibitemOpen
  \bibfield  {author} {\bibinfo {author} {\bibfnamefont {D.~H.}\ \bibnamefont
  {Lee}},\ }\href {\doibase 10.1103/PhysRevLett.103.196804} {\bibfield
  {journal} {\bibinfo  {journal} {Phys. Rev. Lett.}\ }\textbf {\bibinfo
  {volume} {103}},\ \bibinfo {pages} {196804} (\bibinfo {year}
  {2009})}\BibitemShut {NoStop}%
\bibitem [{\citenamefont {Zhang}\ and\ \citenamefont
  {Vishwanath}(2010)}]{Zhang2010}%
  \BibitemOpen
  \bibfield  {author} {\bibinfo {author} {\bibfnamefont {Y.}~\bibnamefont
  {Zhang}}\ and\ \bibinfo {author} {\bibfnamefont {A.}~\bibnamefont
  {Vishwanath}},\ }\href {\doibase 10.1103/PhysRevLett.105.206601} {\bibfield
  {journal} {\bibinfo  {journal} {Phys. Rev. Lett.}\ }\textbf {\bibinfo
  {volume} {105}},\ \bibinfo {pages} {206601} (\bibinfo {year}
  {2010})}\BibitemShut {NoStop}%
\bibitem [{\citenamefont {Foster}\ and\ \citenamefont
  {Nightingale}(2006)}]{book_FosterNightingale}%
  \BibitemOpen
  \bibfield  {author} {\bibinfo {author} {\bibfnamefont {J.}~\bibnamefont
  {Foster}}\ and\ \bibinfo {author} {\bibfnamefont {J.~D.}\ \bibnamefont
  {Nightingale}},\ }\href@noop {} {\emph {\bibinfo {title} {{A Short Course in
  General Relativity}}}},\ \bibinfo {edition} {3rd}\ ed.\ (\bibinfo
  {publisher} {Springer},\ \bibinfo {year} {2006})\BibitemShut {NoStop}%
\bibitem [{\citenamefont {Poisson}(2004)}]{book_Poisson}%
  \BibitemOpen
  \bibfield  {author} {\bibinfo {author} {\bibfnamefont {E.}~\bibnamefont
  {Poisson}},\ }\href@noop {} {\emph {\bibinfo {title} {{A Relativist's
  toolkit}}}}\ (\bibinfo  {publisher} {Cambridge University Press},\ \bibinfo
  {year} {2004})\BibitemShut {NoStop}%
\bibitem [{\citenamefont {Sato}(1996)}]{book_SatoKatsu}%
  \BibitemOpen
  \bibfield  {author} {\bibinfo {author} {\bibfnamefont {K.}~\bibnamefont
  {Sato}},\ }\href@noop {} {\emph {\bibinfo {title} {{The Theory of
  Relativity}}}}\ (\bibinfo  {publisher} {Iwanami Shoten, in Japanese},\
  \bibinfo {year} {1996})\BibitemShut {NoStop}%
\bibitem [{ann({\natexlab{a}})}]{annotation_indices}%
  \BibitemOpen
  \href@noop {} {\bibfield  {journal} {\bibinfo  {journal} {The Cartesian
  indices are raised and lowered by the Kronecker delta, while the curvilinear
  indices by the curvilinear metric tensor. We do not have to make a
  distinction between raised or lowered Cartesian indices}}
 }\BibitemShut {NoStop}%
\bibitem [{ann({\natexlab{b}})}]{annotation_specific}%
  \BibitemOpen
  \href@noop {} {\bibfield  {journal} {\bibinfo  {journal} {This definition of
  spin connections is very specific to the case of curves and surfaces embedded
  in a flat space. See Refs.\cite{book_Carroll,
  book_Nakahara, book_GSW, book_ParkerToms} for a more general
  definition}} }\BibitemShut {NoStop}%
\bibitem [{\citenamefont {Kavalov}\ \emph {et~al.}(1986)\citenamefont
  {Kavalov}, \citenamefont {Kostov},\ and\ \citenamefont
  {Sedrakyan}}]{Kavalov1986}%
  \BibitemOpen
  \bibfield  {author} {\bibinfo {author} {\bibfnamefont {A.~R.}\ \bibnamefont
  {Kavalov}}, \bibinfo {author} {\bibfnamefont {I.~K.}\ \bibnamefont {Kostov}},
  \ and\ \bibinfo {author} {\bibfnamefont {A.~G.}\ \bibnamefont {Sedrakyan}},\
  }\href {\doibase 10.1016/0370-2693(86)90865-8} {\bibfield  {journal}
  {\bibinfo  {journal} {Phys. Lett. B}\ }\textbf {\bibinfo {volume} {175}},\
  \bibinfo {pages} {331} (\bibinfo {year} {1986})}\BibitemShut {NoStop}%
\bibitem [{ann({\natexlab{c}})}]{annotation_David}%
  \BibitemOpen
  \href@noop {} {\bibfield  {journal} {\bibinfo  {journal} {See Eqs.(2.44),
  (2.45) and (7.3) in Ref.\cite{book_David}. Our results
  agree with these equations when we restrict vector fields, on which spin
  connections act, to be tangential to the surface}}
 }\BibitemShut {NoStop}%
\bibitem [{\citenamefont {do~Carmo}(2016)}]{book_doCarmo}%
  \BibitemOpen
  \bibfield  {author} {\bibinfo {author} {\bibfnamefont {M.~P.}\ \bibnamefont
  {do~Carmo}},\ }\href@noop {} {\emph {\bibinfo {title} {{Differential Geometry
  of Curves and Surfaces}}}},\ \bibinfo {edition} {2nd}\ ed.\ (\bibinfo
  {publisher} {Dover Publications},\ \bibinfo {year} {2016})\BibitemShut
  {NoStop}%
\bibitem [{ann({\natexlab{d}})}]{annotation_strain}%
  \BibitemOpen
  \href@noop {} {\bibfield  {journal} {\bibinfo  {journal} {Lattice strains can
  be induced either by externally applied mechanical forces or by the curvature
  of surfaces. Our discussion in this paper corresponds to the latter case, and
  can be extended to the former}} }\BibitemShut {NoStop}%
\bibitem [{\citenamefont {Neto}\ \emph {et~al.}(2009)\citenamefont {Neto},
  \citenamefont {Guinea}, \citenamefont {Peres}, \citenamefont {Novoselov},\
  and\ \citenamefont {Geim}}]{Neto2009}%
  \BibitemOpen
  \bibfield  {author} {\bibinfo {author} {\bibfnamefont {A.~H.~C.}\
  \bibnamefont {Neto}}, \bibinfo {author} {\bibfnamefont {F.}~\bibnamefont
  {Guinea}}, \bibinfo {author} {\bibfnamefont {N.~M.~R.}\ \bibnamefont
  {Peres}}, \bibinfo {author} {\bibfnamefont {K.~S.}\ \bibnamefont
  {Novoselov}}, \ and\ \bibinfo {author} {\bibfnamefont {A.~K.}\ \bibnamefont
  {Geim}},\ }\href {\doibase 10.1103/RevModPhys.81.109} {\bibfield  {journal}
  {\bibinfo  {journal} {Rev. Mod. Phys.}\ }\textbf {\bibinfo {volume} {81}},\
  \bibinfo {pages} {109} (\bibinfo {year} {2009})}\BibitemShut {NoStop}%
\bibitem [{\citenamefont {Levy}\ \emph {et~al.}(2010)\citenamefont {Levy},
  \citenamefont {Burke}, \citenamefont {Meaker}, \citenamefont {Panlasigui},
  \citenamefont {Zettl}, \citenamefont {Guinea}, \citenamefont {Neto},\ and\
  \citenamefont {Crommie}}]{Levy2010}%
  \BibitemOpen
  \bibfield  {author} {\bibinfo {author} {\bibfnamefont {N.}~\bibnamefont
  {Levy}}, \bibinfo {author} {\bibfnamefont {S.~A.}\ \bibnamefont {Burke}},
  \bibinfo {author} {\bibfnamefont {K.~L.}\ \bibnamefont {Meaker}}, \bibinfo
  {author} {\bibfnamefont {M.}~\bibnamefont {Panlasigui}}, \bibinfo {author}
  {\bibfnamefont {A.}~\bibnamefont {Zettl}}, \bibinfo {author} {\bibfnamefont
  {F.}~\bibnamefont {Guinea}}, \bibinfo {author} {\bibfnamefont {A.~H.~C.}\
  \bibnamefont {Neto}}, \ and\ \bibinfo {author} {\bibfnamefont {M.~F.}\
  \bibnamefont {Crommie}},\ }\href {\doibase 10.1126/science.1191700}
  {\bibfield  {journal} {\bibinfo  {journal} {Science}\ }\textbf {\bibinfo
  {volume} {329}},\ \bibinfo {pages} {544} (\bibinfo {year}
  {2010})}\BibitemShut {NoStop}%
\bibitem [{\citenamefont {Vozmediano}\ \emph {et~al.}(2010)\citenamefont
  {Vozmediano}, \citenamefont {Katsnelson},\ and\ \citenamefont
  {Guinea}}]{Vozmediano2010}%
  \BibitemOpen
  \bibfield  {author} {\bibinfo {author} {\bibfnamefont {M.~A.}\ \bibnamefont
  {Vozmediano}}, \bibinfo {author} {\bibfnamefont {M.~I.}\ \bibnamefont
  {Katsnelson}}, \ and\ \bibinfo {author} {\bibfnamefont {F.}~\bibnamefont
  {Guinea}},\ }\href {\doibase 10.1016/j.physrep.2010.07.003} {\bibfield
  {journal} {\bibinfo  {journal} {Phys. Rep.}\ }\textbf {\bibinfo {volume}
  {496}},\ \bibinfo {pages} {109} (\bibinfo {year} {2010})}\BibitemShut
  {NoStop}%
\bibitem [{\citenamefont {Katsnelson}(2012)}]{book_Katsnelson}%
  \BibitemOpen
  \bibfield  {author} {\bibinfo {author} {\bibfnamefont {M.~I.}\ \bibnamefont
  {Katsnelson}},\ }\href@noop {} {\emph {\bibinfo {title} {{Graphene: Carbon in
  Two Dimensions}}}}\ (\bibinfo  {publisher} {Cambridge University Press},\
  \bibinfo {year} {2012})\BibitemShut {NoStop}%
\bibitem [{\citenamefont {Amorim}\ \emph {et~al.}(2016)\citenamefont {Amorim},
  \citenamefont {Cortijo}, \citenamefont {{De Juan}}, \citenamefont {Grushin},
  \citenamefont {Guinea}, \citenamefont {Guti{\'{e}}rrez-Rubio}, \citenamefont
  {Ochoa}, \citenamefont {Parente}, \citenamefont {Rold{\'{a}}n}, \citenamefont
  {San-Jose}, \citenamefont {Schiefele}, \citenamefont {Sturla},\ and\
  \citenamefont {Vozmediano}}]{Amorim2016}%
  \BibitemOpen
  \bibfield  {author} {\bibinfo {author} {\bibfnamefont {B.}~\bibnamefont
  {Amorim}}, \bibinfo {author} {\bibfnamefont {A.}~\bibnamefont {Cortijo}},
  \bibinfo {author} {\bibfnamefont {F.}~\bibnamefont {{De Juan}}}, \bibinfo
  {author} {\bibfnamefont {A.~G.}\ \bibnamefont {Grushin}}, \bibinfo {author}
  {\bibfnamefont {F.}~\bibnamefont {Guinea}}, \bibinfo {author} {\bibfnamefont
  {A.}~\bibnamefont {Guti{\'{e}}rrez-Rubio}}, \bibinfo {author} {\bibfnamefont
  {H.}~\bibnamefont {Ochoa}}, \bibinfo {author} {\bibfnamefont
  {V.}~\bibnamefont {Parente}}, \bibinfo {author} {\bibfnamefont
  {R.}~\bibnamefont {Rold{\'{a}}n}}, \bibinfo {author} {\bibfnamefont
  {P.}~\bibnamefont {San-Jose}}, \bibinfo {author} {\bibfnamefont
  {J.}~\bibnamefont {Schiefele}}, \bibinfo {author} {\bibfnamefont
  {M.}~\bibnamefont {Sturla}}, \ and\ \bibinfo {author} {\bibfnamefont {M.~A.}\
  \bibnamefont {Vozmediano}},\ }\href {\doibase 10.1016/j.physrep.2015.12.006}
  {\bibfield  {journal} {\bibinfo  {journal} {Phys. Rep.}\ }\textbf {\bibinfo
  {volume} {617}},\ \bibinfo {pages} {1} (\bibinfo {year} {2016})}\BibitemShut
  {NoStop}%
\bibitem [{\citenamefont {Yin}\ \emph {et~al.}(2017)\citenamefont {Yin},
  \citenamefont {Bai}, \citenamefont {Wang}, \citenamefont {Li}, \citenamefont
  {Zhang},\ and\ \citenamefont {He}}]{Yi2017}%
  \BibitemOpen
  \bibfield  {author} {\bibinfo {author} {\bibfnamefont {L.~J.}\ \bibnamefont
  {Yin}}, \bibinfo {author} {\bibfnamefont {K.~K.}\ \bibnamefont {Bai}},
  \bibinfo {author} {\bibfnamefont {W.~X.}\ \bibnamefont {Wang}}, \bibinfo
  {author} {\bibfnamefont {S.~Y.}\ \bibnamefont {Li}}, \bibinfo {author}
  {\bibfnamefont {Y.}~\bibnamefont {Zhang}}, \ and\ \bibinfo {author}
  {\bibfnamefont {L.}~\bibnamefont {He}},\ }\href {\doibase
  10.1007/s11467-017-0655-0} {\bibfield  {journal} {\bibinfo  {journal} {Front.
  Phys.}\ }\textbf {\bibinfo {volume} {12}},\ \bibinfo {pages} {127208}
  (\bibinfo {year} {2017})}\BibitemShut {NoStop}%
\bibitem [{\citenamefont {Naumis}\ \emph {et~al.}(2017)\citenamefont {Naumis},
  \citenamefont {Barraza-Lopez}, \citenamefont {Oliva-Leyva},\ and\
  \citenamefont {Terrones}}]{Naumis2017}%
  \BibitemOpen
  \bibfield  {author} {\bibinfo {author} {\bibfnamefont {G.~G.}\ \bibnamefont
  {Naumis}}, \bibinfo {author} {\bibfnamefont {S.}~\bibnamefont
  {Barraza-Lopez}}, \bibinfo {author} {\bibfnamefont {M.}~\bibnamefont
  {Oliva-Leyva}}, \ and\ \bibinfo {author} {\bibfnamefont {H.}~\bibnamefont
  {Terrones}},\ }\href {\doibase 10.1088/1361-6633/aa74ef} {\bibfield
  {journal} {\bibinfo  {journal} {Rep. Prog. Phys.}\ }\textbf {\bibinfo
  {volume} {80}},\ \bibinfo {pages} {096501} (\bibinfo {year}
  {2017})}\BibitemShut {NoStop}%
\bibitem [{\citenamefont {Cazalilla}\ \emph {et~al.}(2014)\citenamefont
  {Cazalilla}, \citenamefont {Ochoa},\ and\ \citenamefont
  {Guinea}}]{Cazalilla2014}%
  \BibitemOpen
  \bibfield  {author} {\bibinfo {author} {\bibfnamefont {M.~A.}\ \bibnamefont
  {Cazalilla}}, \bibinfo {author} {\bibfnamefont {H.}~\bibnamefont {Ochoa}}, \
  and\ \bibinfo {author} {\bibfnamefont {F.}~\bibnamefont {Guinea}},\ }\href
  {\doibase 10.1103/PhysRevLett.113.077201} {\bibfield  {journal} {\bibinfo
  {journal} {Phys. Rev. Lett.}\ }\textbf {\bibinfo {volume} {113}},\ \bibinfo
  {pages} {077201} (\bibinfo {year} {2014})}\BibitemShut {NoStop}%
\bibitem [{\citenamefont {Cortijo}\ \emph {et~al.}(2015)\citenamefont
  {Cortijo}, \citenamefont {Ferreir{\'{o}}s}, \citenamefont {Landsteiner},\
  and\ \citenamefont {Vozmediano}}]{Cortijo2015}%
  \BibitemOpen
  \bibfield  {author} {\bibinfo {author} {\bibfnamefont {A.}~\bibnamefont
  {Cortijo}}, \bibinfo {author} {\bibfnamefont {Y.}~\bibnamefont
  {Ferreir{\'{o}}s}}, \bibinfo {author} {\bibfnamefont {K.}~\bibnamefont
  {Landsteiner}}, \ and\ \bibinfo {author} {\bibfnamefont {M.~A.}\ \bibnamefont
  {Vozmediano}},\ }\href {\doibase 10.1103/PhysRevLett.115.177202} {\bibfield
  {journal} {\bibinfo  {journal} {Phys. Rev. Lett.}\ }\textbf {\bibinfo
  {volume} {115}},\ \bibinfo {pages} {177202} (\bibinfo {year}
  {2015})}\BibitemShut {NoStop}%
\bibitem [{\citenamefont {Shapourian}\ \emph {et~al.}(2015)\citenamefont
  {Shapourian}, \citenamefont {Hughes},\ and\ \citenamefont
  {Ryu}}]{Shapourian2015}%
  \BibitemOpen
  \bibfield  {author} {\bibinfo {author} {\bibfnamefont {H.}~\bibnamefont
  {Shapourian}}, \bibinfo {author} {\bibfnamefont {T.~L.}\ \bibnamefont
  {Hughes}}, \ and\ \bibinfo {author} {\bibfnamefont {S.}~\bibnamefont {Ryu}},\
  }\href {\doibase 10.1103/PhysRevB.92.165131} {\bibfield  {journal} {\bibinfo
  {journal} {Phys. Rev. B}\ }\textbf {\bibinfo {volume} {92}},\ \bibinfo
  {pages} {165131} (\bibinfo {year} {2015})}\BibitemShut {NoStop}%
\bibitem [{ann({\natexlab{e}})}]{annotation_nongeometric_indices}%
  \BibitemOpen
  \href@noop {} {\bibfield  {journal} {\bibinfo  {journal} {We note that in
  some cases indices attached to spinors are non-geometrical. For example, the
  so-called isospin degrees of freedom in nuclear physics express a proton as
  the up-state and a neutron as the down-state. Spin connections do not act on
  such a kind of indices}} }\BibitemShut {NoStop}%
\bibitem [{\citenamefont {Rashba}(2003)}]{Rashba2003}%
  \BibitemOpen
  \bibfield  {author} {\bibinfo {author} {\bibfnamefont {E.~I.}\ \bibnamefont
  {Rashba}},\ }\href {\doibase 10.1103/PhysRevB.68.241315} {\bibfield
  {journal} {\bibinfo  {journal} {Phys. Rev. B}\ }\textbf {\bibinfo {volume}
  {68}},\ \bibinfo {pages} {241315(R)} (\bibinfo {year} {2003})}\BibitemShut
  {NoStop}%
\bibitem [{\citenamefont {Katsnelson}\ \emph {et~al.}(2010)\citenamefont
  {Katsnelson}, \citenamefont {Kvashnin}, \citenamefont {Mazurenko},\ and\
  \citenamefont {Lichtenstein}}]{Katsnelson2010}%
  \BibitemOpen
  \bibfield  {author} {\bibinfo {author} {\bibfnamefont {M.~I.}\ \bibnamefont
  {Katsnelson}}, \bibinfo {author} {\bibfnamefont {Y.~O.}\ \bibnamefont
  {Kvashnin}}, \bibinfo {author} {\bibfnamefont {V.~V.}\ \bibnamefont
  {Mazurenko}}, \ and\ \bibinfo {author} {\bibfnamefont {A.~I.}\ \bibnamefont
  {Lichtenstein}},\ }\href {\doibase 10.1103/PhysRevB.82.100403} {\bibfield
  {journal} {\bibinfo  {journal} {Phys. Rev. B}\ }\textbf {\bibinfo {volume}
  {82}},\ \bibinfo {pages} {100403(R)} (\bibinfo {year} {2010})}\BibitemShut
  {NoStop}%
\bibitem [{\citenamefont {Kikuchi}\ \emph {et~al.}(2016)\citenamefont
  {Kikuchi}, \citenamefont {Koretsune}, \citenamefont {Arita},\ and\
  \citenamefont {Tatara}}]{Kikuchi2016}%
  \BibitemOpen
  \bibfield  {author} {\bibinfo {author} {\bibfnamefont {T.}~\bibnamefont
  {Kikuchi}}, \bibinfo {author} {\bibfnamefont {T.}~\bibnamefont {Koretsune}},
  \bibinfo {author} {\bibfnamefont {R.}~\bibnamefont {Arita}}, \ and\ \bibinfo
  {author} {\bibfnamefont {G.}~\bibnamefont {Tatara}},\ }\href {\doibase
  10.1103/PhysRevLett.116.247201} {\bibfield  {journal} {\bibinfo  {journal}
  {Phys. Rev. Lett.}\ }\textbf {\bibinfo {volume} {116}},\ \bibinfo {pages}
  {247201} (\bibinfo {year} {2016})}\BibitemShut {NoStop}%
\bibitem [{\citenamefont {Koretsune}\ \emph {et~al.}(2018)\citenamefont
  {Koretsune}, \citenamefont {Kikuchi},\ and\ \citenamefont
  {Arita}}]{Koretsune2018}%
  \BibitemOpen
  \bibfield  {author} {\bibinfo {author} {\bibfnamefont {T.}~\bibnamefont
  {Koretsune}}, \bibinfo {author} {\bibfnamefont {T.}~\bibnamefont {Kikuchi}},
  \ and\ \bibinfo {author} {\bibfnamefont {R.}~\bibnamefont {Arita}},\ }\href
  {\doibase 10.7566/JPSJ.87.041011} {\bibfield  {journal} {\bibinfo  {journal}
  {J. Phys. Soc. Jpn.}\ }\textbf {\bibinfo {volume} {87}},\ \bibinfo {pages}
  {041011} (\bibinfo {year} {2018})}\BibitemShut {NoStop}%
\bibitem [{\citenamefont {Yosida}(1998)}]{book_YosidaEnglish}%
  \BibitemOpen
  \bibfield  {author} {\bibinfo {author} {\bibfnamefont {K.}~\bibnamefont
  {Yosida}},\ }\href@noop {} {\emph {\bibinfo {title} {{Theory of
  Magnetism}}}}\ (\bibinfo  {publisher} {Springer},\ \bibinfo {year}
  {1998})\BibitemShut {NoStop}%
\bibitem [{\citenamefont {Skomski}(2008)}]{book_Skomski}%
  \BibitemOpen
  \bibfield  {author} {\bibinfo {author} {\bibfnamefont {R.}~\bibnamefont
  {Skomski}},\ }\href@noop {} {\emph {\bibinfo {title} {{Simple Models of
  Magnetism}}}}\ (\bibinfo  {publisher} {Oxford University Press},\ \bibinfo
  {year} {2008})\BibitemShut {NoStop}%
\bibitem [{\citenamefont {Kanamori}(1969)}]{book_Kanamori}%
  \BibitemOpen
  \bibfield  {author} {\bibinfo {author} {\bibfnamefont {J.}~\bibnamefont
  {Kanamori}},\ }\href@noop {} {\emph {\bibinfo {title} {{Magnetism}}}}\
  (\bibinfo  {publisher} {Baifukan, in Japanese},\ \bibinfo {year}
  {1969})\BibitemShut {NoStop}%
\bibitem [{\citenamefont {Yosida}(1991)}]{book_Yosida}%
  \BibitemOpen
  \bibfield  {author} {\bibinfo {author} {\bibfnamefont {K.}~\bibnamefont
  {Yosida}},\ }\href@noop {} {\emph {\bibinfo {title} {{Magnetism}}}}\
  (\bibinfo  {publisher} {Iwanami Shoten, in Japanese},\ \bibinfo {year}
  {1991})\BibitemShut {NoStop}%
\bibitem [{\citenamefont {Matsuda}(1993)}]{book_Matsuda}%
  \BibitemOpen
  \bibfield  {author} {\bibinfo {author} {\bibfnamefont {S.}~\bibnamefont
  {Matsuda}},\ }\href@noop {} {\emph {\bibinfo {title} {{Mechanics}}}},\ edited
  by\ \bibinfo {editor} {\bibfnamefont {J.}~\bibnamefont {Maki}}, \bibinfo
  {editor} {\bibfnamefont {Y.}~\bibnamefont {Nagaoka}}, \ and\ \bibinfo
  {editor} {\bibfnamefont {Y.}~\bibnamefont {Ohtsuki}}\ (\bibinfo  {publisher}
  {Maruzen, in Japanese},\ \bibinfo {year} {1993})\BibitemShut {NoStop}%
\bibitem [{\citenamefont {Coury}\ \emph {et~al.}(2016)\citenamefont {Coury},
  \citenamefont {Dudarev}, \citenamefont {Foulkes}, \citenamefont {Horsfield},
  \citenamefont {Ma},\ and\ \citenamefont {Spencer}}]{Coury2016}%
  \BibitemOpen
  \bibfield  {author} {\bibinfo {author} {\bibfnamefont {M.~E.}\ \bibnamefont
  {Coury}}, \bibinfo {author} {\bibfnamefont {S.~L.}\ \bibnamefont {Dudarev}},
  \bibinfo {author} {\bibfnamefont {W.~M.}\ \bibnamefont {Foulkes}}, \bibinfo
  {author} {\bibfnamefont {A.~P.}\ \bibnamefont {Horsfield}}, \bibinfo {author}
  {\bibfnamefont {P.~W.}\ \bibnamefont {Ma}}, \ and\ \bibinfo {author}
  {\bibfnamefont {J.~S.}\ \bibnamefont {Spencer}},\ }\href {\doibase
  10.1103/PhysRevB.93.075101} {\bibfield  {journal} {\bibinfo  {journal} {Phys.
  Rev. B}\ }\textbf {\bibinfo {volume} {93}},\ \bibinfo {pages} {075101}
  (\bibinfo {year} {2016})}\BibitemShut {NoStop}%
\bibitem [{\citenamefont {Wang}\ \emph {et~al.}(2012)\citenamefont {Wang},
  \citenamefont {Kalantar-Zadeh}, \citenamefont {Kis}, \citenamefont
  {Coleman},\ and\ \citenamefont {Strano}}]{Wang2012}%
  \BibitemOpen
  \bibfield  {author} {\bibinfo {author} {\bibfnamefont {Q.~H.}\ \bibnamefont
  {Wang}}, \bibinfo {author} {\bibfnamefont {K.}~\bibnamefont
  {Kalantar-Zadeh}}, \bibinfo {author} {\bibfnamefont {A.}~\bibnamefont {Kis}},
  \bibinfo {author} {\bibfnamefont {J.~N.}\ \bibnamefont {Coleman}}, \ and\
  \bibinfo {author} {\bibfnamefont {M.~S.}\ \bibnamefont {Strano}},\ }\href
  {\doibase 10.1038/nnano.2012.193} {\bibfield  {journal} {\bibinfo  {journal}
  {Nat. Nanotech.}\ }\textbf {\bibinfo {volume} {7}},\ \bibinfo {pages} {699}
  (\bibinfo {year} {2012})}\BibitemShut {NoStop}%
\bibitem [{\citenamefont {Choi}\ \emph {et~al.}(2017)\citenamefont {Choi},
  \citenamefont {Choudhary}, \citenamefont {Han}, \citenamefont {Park},
  \citenamefont {Akinwande},\ and\ \citenamefont {Lee}}]{Choi2017}%
  \BibitemOpen
  \bibfield  {author} {\bibinfo {author} {\bibfnamefont {W.}~\bibnamefont
  {Choi}}, \bibinfo {author} {\bibfnamefont {N.}~\bibnamefont {Choudhary}},
  \bibinfo {author} {\bibfnamefont {G.~H.}\ \bibnamefont {Han}}, \bibinfo
  {author} {\bibfnamefont {J.}~\bibnamefont {Park}}, \bibinfo {author}
  {\bibfnamefont {D.}~\bibnamefont {Akinwande}}, \ and\ \bibinfo {author}
  {\bibfnamefont {Y.~H.}\ \bibnamefont {Lee}},\ }\href {\doibase
  10.1016/j.mattod.2016.10.002} {\bibfield  {journal} {\bibinfo  {journal}
  {Mater. Today}\ }\textbf {\bibinfo {volume} {20}},\ \bibinfo {pages} {116}
  (\bibinfo {year} {2017})}\BibitemShut {NoStop}%
\bibitem [{\citenamefont {Qin}\ \emph {et~al.}(2017)\citenamefont {Qin},
  \citenamefont {Shi}, \citenamefont {Ideue}, \citenamefont {Yoshida},
  \citenamefont {Zak}, \citenamefont {Tenne}, \citenamefont {Kikitsu},
  \citenamefont {Inoue}, \citenamefont {Hashizume},\ and\ \citenamefont
  {Iwasa}}]{Qin2017}%
  \BibitemOpen
  \bibfield  {author} {\bibinfo {author} {\bibfnamefont {F.}~\bibnamefont
  {Qin}}, \bibinfo {author} {\bibfnamefont {W.}~\bibnamefont {Shi}}, \bibinfo
  {author} {\bibfnamefont {T.}~\bibnamefont {Ideue}}, \bibinfo {author}
  {\bibfnamefont {M.}~\bibnamefont {Yoshida}}, \bibinfo {author} {\bibfnamefont
  {A.}~\bibnamefont {Zak}}, \bibinfo {author} {\bibfnamefont {R.}~\bibnamefont
  {Tenne}}, \bibinfo {author} {\bibfnamefont {T.}~\bibnamefont {Kikitsu}},
  \bibinfo {author} {\bibfnamefont {D.}~\bibnamefont {Inoue}}, \bibinfo
  {author} {\bibfnamefont {D.}~\bibnamefont {Hashizume}}, \ and\ \bibinfo
  {author} {\bibfnamefont {Y.}~\bibnamefont {Iwasa}},\ }\href {\doibase
  10.1038/ncomms14465} {\bibfield  {journal} {\bibinfo  {journal} {Nat.
  Commun.}\ }\textbf {\bibinfo {volume} {8}},\ \bibinfo {pages} {14465}
  (\bibinfo {year} {2017})}\BibitemShut {NoStop}%
\bibitem [{\citenamefont {Manzeli}\ \emph {et~al.}(2017)\citenamefont
  {Manzeli}, \citenamefont {Ovchinnikov}, \citenamefont {Pasquier},
  \citenamefont {Yazyev},\ and\ \citenamefont {Kis}}]{Manzeli2017}%
  \BibitemOpen
  \bibfield  {author} {\bibinfo {author} {\bibfnamefont {S.}~\bibnamefont
  {Manzeli}}, \bibinfo {author} {\bibfnamefont {D.}~\bibnamefont
  {Ovchinnikov}}, \bibinfo {author} {\bibfnamefont {D.}~\bibnamefont
  {Pasquier}}, \bibinfo {author} {\bibfnamefont {O.~V.}\ \bibnamefont
  {Yazyev}}, \ and\ \bibinfo {author} {\bibfnamefont {A.}~\bibnamefont {Kis}},\
  }\href {\doibase 10.1038/natrevmats.2017.33} {\bibfield  {journal} {\bibinfo
  {journal} {Nat. Rev. Mater.}\ }\textbf {\bibinfo {volume} {2}},\ \bibinfo
  {pages} {17033} (\bibinfo {year} {2017})}\BibitemShut {NoStop}%
\bibitem [{\citenamefont {Dong}\ and\ \citenamefont
  {Kuljanishvili}(2017)}]{Dong2017}%
  \BibitemOpen
  \bibfield  {author} {\bibinfo {author} {\bibfnamefont {R.}~\bibnamefont
  {Dong}}\ and\ \bibinfo {author} {\bibfnamefont {I.}~\bibnamefont
  {Kuljanishvili}},\ }\href {\doibase 10.1116/1.4982736} {\bibfield  {journal}
  {\bibinfo  {journal} {J. Vac. Sci. Technol. B: Nanotechnol. Microelectron.}\
  }\textbf {\bibinfo {volume} {35}},\ \bibinfo {pages} {030803} (\bibinfo
  {year} {2017})}\BibitemShut {NoStop}%
\bibitem [{\citenamefont {Bernevig}\ \emph {et~al.}(2005)\citenamefont
  {Bernevig}, \citenamefont {Hughes},\ and\ \citenamefont
  {Zhang}}]{Bernevig2005}%
  \BibitemOpen
  \bibfield  {author} {\bibinfo {author} {\bibfnamefont {B.~A.}\ \bibnamefont
  {Bernevig}}, \bibinfo {author} {\bibfnamefont {T.~L.}\ \bibnamefont
  {Hughes}}, \ and\ \bibinfo {author} {\bibfnamefont {S.~C.}\ \bibnamefont
  {Zhang}},\ }\href {\doibase 10.1103/PhysRevLett.95.066601} {\bibfield
  {journal} {\bibinfo  {journal} {Phys. Rev. Lett.}\ }\textbf {\bibinfo
  {volume} {95}},\ \bibinfo {pages} {066601} (\bibinfo {year}
  {2005})}\BibitemShut {NoStop}%
\bibitem [{\citenamefont {Kontani}\ \emph {et~al.}(2008)\citenamefont
  {Kontani}, \citenamefont {Tanaka}, \citenamefont {Hirashima}, \citenamefont
  {Yamada},\ and\ \citenamefont {Inoue}}]{Kontani2008PRL}%
  \BibitemOpen
  \bibfield  {author} {\bibinfo {author} {\bibfnamefont {H.}~\bibnamefont
  {Kontani}}, \bibinfo {author} {\bibfnamefont {T.}~\bibnamefont {Tanaka}},
  \bibinfo {author} {\bibfnamefont {D.~S.}\ \bibnamefont {Hirashima}}, \bibinfo
  {author} {\bibfnamefont {K.}~\bibnamefont {Yamada}}, \ and\ \bibinfo {author}
  {\bibfnamefont {J.}~\bibnamefont {Inoue}},\ }\href {\doibase
  10.1103/PhysRevLett.100.096601} {\bibfield  {journal} {\bibinfo  {journal}
  {Phys. Rev. Lett}\ }\textbf {\bibinfo {volume} {100}},\ \bibinfo {pages}
  {096601} (\bibinfo {year} {2008})}\BibitemShut {NoStop}%
\bibitem [{\citenamefont {Tanaka}\ \emph {et~al.}(2008)\citenamefont {Tanaka},
  \citenamefont {Kontani}, \citenamefont {Naito}, \citenamefont {Naito},
  \citenamefont {Hirashima}, \citenamefont {Yamada},\ and\ \citenamefont
  {Inoue}}]{Tanaka2008}%
  \BibitemOpen
  \bibfield  {author} {\bibinfo {author} {\bibfnamefont {T.}~\bibnamefont
  {Tanaka}}, \bibinfo {author} {\bibfnamefont {H.}~\bibnamefont {Kontani}},
  \bibinfo {author} {\bibfnamefont {M.}~\bibnamefont {Naito}}, \bibinfo
  {author} {\bibfnamefont {T.}~\bibnamefont {Naito}}, \bibinfo {author}
  {\bibfnamefont {D.~S.}\ \bibnamefont {Hirashima}}, \bibinfo {author}
  {\bibfnamefont {K.}~\bibnamefont {Yamada}}, \ and\ \bibinfo {author}
  {\bibfnamefont {J.}~\bibnamefont {Inoue}},\ }\href {\doibase
  10.1103/PhysRevB.77.165117} {\bibfield  {journal} {\bibinfo  {journal} {Phys.
  Rev. B}\ }\textbf {\bibinfo {volume} {77}},\ \bibinfo {pages} {165117}
  (\bibinfo {year} {2008})}\BibitemShut {NoStop}%
\bibitem [{\citenamefont {Kontani}\ \emph {et~al.}(2009)\citenamefont
  {Kontani}, \citenamefont {Tanaka}, \citenamefont {Hirashima}, \citenamefont
  {Yamada},\ and\ \citenamefont {Inoue}}]{Kontani2009PRL}%
  \BibitemOpen
  \bibfield  {author} {\bibinfo {author} {\bibfnamefont {H.}~\bibnamefont
  {Kontani}}, \bibinfo {author} {\bibfnamefont {T.}~\bibnamefont {Tanaka}},
  \bibinfo {author} {\bibfnamefont {D.~S.}\ \bibnamefont {Hirashima}}, \bibinfo
  {author} {\bibfnamefont {K.}~\bibnamefont {Yamada}}, \ and\ \bibinfo {author}
  {\bibfnamefont {J.}~\bibnamefont {Inoue}},\ }\href {\doibase
  10.1103/PhysRevLett.102.016601} {\bibfield  {journal} {\bibinfo  {journal}
  {Phys. Rev. Lett.}\ }\textbf {\bibinfo {volume} {102}},\ \bibinfo {pages}
  {016601} (\bibinfo {year} {2009})}\BibitemShut {NoStop}%
\bibitem [{\citenamefont {Bernevig}\ and\ \citenamefont
  {Zhang}(2006)}]{Bernevig2006}%
  \BibitemOpen
  \bibfield  {author} {\bibinfo {author} {\bibfnamefont {B.~A.}\ \bibnamefont
  {Bernevig}}\ and\ \bibinfo {author} {\bibfnamefont {S.-C.}\ \bibnamefont
  {Zhang}},\ }\href {\doibase 10.1103/PhysRevLett.96.106802} {\bibfield
  {journal} {\bibinfo  {journal} {Phys. Rev. Lett.}\ }\textbf {\bibinfo
  {volume} {96}},\ \bibinfo {pages} {106802} (\bibinfo {year}
  {2006})}\BibitemShut {NoStop}%
\bibitem [{\citenamefont {Matsuo}\ \emph {et~al.}(2011)\citenamefont {Matsuo},
  \citenamefont {Ieda}, \citenamefont {Saitoh},\ and\ \citenamefont
  {Maekawa}}]{Matsuo2011}%
  \BibitemOpen
  \bibfield  {author} {\bibinfo {author} {\bibfnamefont {M.}~\bibnamefont
  {Matsuo}}, \bibinfo {author} {\bibfnamefont {J.}~\bibnamefont {Ieda}},
  \bibinfo {author} {\bibfnamefont {E.}~\bibnamefont {Saitoh}}, \ and\ \bibinfo
  {author} {\bibfnamefont {S.}~\bibnamefont {Maekawa}},\ }\href {\doibase
  10.1103/PhysRevLett.106.076601} {\bibfield  {journal} {\bibinfo  {journal}
  {Phys. Rev. Lett.}\ }\textbf {\bibinfo {volume} {106}},\ \bibinfo {pages}
  {076601} (\bibinfo {year} {2011})}\BibitemShut {NoStop}%
\bibitem [{\citenamefont {Matsuo}\ \emph {et~al.}(2013)\citenamefont {Matsuo},
  \citenamefont {Ieda}, \citenamefont {Harii}, \citenamefont {Saitoh},\ and\
  \citenamefont {Maekawa}}]{Matsuo2013}%
  \BibitemOpen
  \bibfield  {author} {\bibinfo {author} {\bibfnamefont {M.}~\bibnamefont
  {Matsuo}}, \bibinfo {author} {\bibfnamefont {J.}~\bibnamefont {Ieda}},
  \bibinfo {author} {\bibfnamefont {K.}~\bibnamefont {Harii}}, \bibinfo
  {author} {\bibfnamefont {E.}~\bibnamefont {Saitoh}}, \ and\ \bibinfo {author}
  {\bibfnamefont {S.}~\bibnamefont {Maekawa}},\ }\href {\doibase
  10.1103/PhysRevB.87.180402} {\bibfield  {journal} {\bibinfo  {journal} {Phys.
  Rev. B}\ }\textbf {\bibinfo {volume} {87}},\ \bibinfo {pages} {180402(R)}
  (\bibinfo {year} {2013})}\BibitemShut {NoStop}%
\bibitem [{\citenamefont {Qian}\ \emph {et~al.}(2014)\citenamefont {Qian},
  \citenamefont {Liu}, \citenamefont {Fu},\ and\ \citenamefont
  {Li}}]{Qian2014}%
  \BibitemOpen
  \bibfield  {author} {\bibinfo {author} {\bibfnamefont {X.}~\bibnamefont
  {Qian}}, \bibinfo {author} {\bibfnamefont {J.}~\bibnamefont {Liu}}, \bibinfo
  {author} {\bibfnamefont {L.}~\bibnamefont {Fu}}, \ and\ \bibinfo {author}
  {\bibfnamefont {J.}~\bibnamefont {Li}},\ }\href {\doibase
  10.1126/science.1256815} {\bibfield  {journal} {\bibinfo  {journal}
  {Science}\ }\textbf {\bibinfo {volume} {346}},\ \bibinfo {pages} {1344}
  (\bibinfo {year} {2014})}\BibitemShut {NoStop}%
\bibitem [{\citenamefont {Matsuo}\ \emph {et~al.}(2017)\citenamefont {Matsuo},
  \citenamefont {Saitoh},\ and\ \citenamefont {Maekawa}}]{Matsuo2017}%
  \BibitemOpen
  \bibfield  {author} {\bibinfo {author} {\bibfnamefont {M.}~\bibnamefont
  {Matsuo}}, \bibinfo {author} {\bibfnamefont {E.}~\bibnamefont {Saitoh}}, \
  and\ \bibinfo {author} {\bibfnamefont {S.}~\bibnamefont {Maekawa}},\ }\href
  {\doibase 10.7566/JPSJ.86.011011} {\bibfield  {journal} {\bibinfo  {journal}
  {J. Phys. Soc. Jpn.}\ }\textbf {\bibinfo {volume} {86}},\ \bibinfo {pages}
  {011011} (\bibinfo {year} {2017})}\BibitemShut {NoStop}%
\bibitem [{\citenamefont {Kobayashi}\ \emph {et~al.}(2017)\citenamefont
  {Kobayashi}, \citenamefont {Yoshikawa}, \citenamefont {Matsuo}, \citenamefont
  {Iguchi}, \citenamefont {Maekawa}, \citenamefont {Saitoh},\ and\
  \citenamefont {Nozaki}}]{Kobayashi2017}%
  \BibitemOpen
  \bibfield  {author} {\bibinfo {author} {\bibfnamefont {D.}~\bibnamefont
  {Kobayashi}}, \bibinfo {author} {\bibfnamefont {T.}~\bibnamefont
  {Yoshikawa}}, \bibinfo {author} {\bibfnamefont {M.}~\bibnamefont {Matsuo}},
  \bibinfo {author} {\bibfnamefont {R.}~\bibnamefont {Iguchi}}, \bibinfo
  {author} {\bibfnamefont {S.}~\bibnamefont {Maekawa}}, \bibinfo {author}
  {\bibfnamefont {E.}~\bibnamefont {Saitoh}}, \ and\ \bibinfo {author}
  {\bibfnamefont {Y.}~\bibnamefont {Nozaki}},\ }\href {\doibase
  10.1103/PhysRevLett.119.077202} {\bibfield  {journal} {\bibinfo  {journal}
  {Phys. Rev. Lett.}\ }\textbf {\bibinfo {volume} {119}},\ \bibinfo {pages}
  {077202} (\bibinfo {year} {2017})}\BibitemShut {NoStop}%
\bibitem [{\citenamefont {Gaididei}\ \emph {et~al.}(2014)\citenamefont
  {Gaididei}, \citenamefont {Kravchuk},\ and\ \citenamefont
  {Sheka}}]{Gaididei2014}%
  \BibitemOpen
  \bibfield  {author} {\bibinfo {author} {\bibfnamefont {Y.}~\bibnamefont
  {Gaididei}}, \bibinfo {author} {\bibfnamefont {V.~P.}\ \bibnamefont
  {Kravchuk}}, \ and\ \bibinfo {author} {\bibfnamefont {D.~D.}\ \bibnamefont
  {Sheka}},\ }\href {\doibase 10.1103/PhysRevLett.112.257203} {\bibfield
  {journal} {\bibinfo  {journal} {Phys. Rev. Lett.}\ }\textbf {\bibinfo
  {volume} {112}},\ \bibinfo {pages} {257203} (\bibinfo {year}
  {2014})}\BibitemShut {NoStop}%
\bibitem [{\citenamefont {Pylypovskyi}\ \emph {et~al.}(2015)\citenamefont
  {Pylypovskyi}, \citenamefont {Kravchuk}, \citenamefont {Sheka}, \citenamefont
  {Makarov}, \citenamefont {Schmidt},\ and\ \citenamefont
  {Gaididei}}]{Pylypovskyi2015}%
  \BibitemOpen
  \bibfield  {author} {\bibinfo {author} {\bibfnamefont {O.~V.}\ \bibnamefont
  {Pylypovskyi}}, \bibinfo {author} {\bibfnamefont {V.~P.}\ \bibnamefont
  {Kravchuk}}, \bibinfo {author} {\bibfnamefont {D.~D.}\ \bibnamefont {Sheka}},
  \bibinfo {author} {\bibfnamefont {D.}~\bibnamefont {Makarov}}, \bibinfo
  {author} {\bibfnamefont {O.~G.}\ \bibnamefont {Schmidt}}, \ and\ \bibinfo
  {author} {\bibfnamefont {Y.}~\bibnamefont {Gaididei}},\ }\href {\doibase
  10.1103/PhysRevLett.114.197204} {\bibfield  {journal} {\bibinfo  {journal}
  {Phys. Rev. Lett.}\ }\textbf {\bibinfo {volume} {114}},\ \bibinfo {pages}
  {197204} (\bibinfo {year} {2015})}\BibitemShut {NoStop}%
\bibitem [{\citenamefont {Sheka}\ \emph {et~al.}(2015)\citenamefont {Sheka},
  \citenamefont {Kravchuk},\ and\ \citenamefont {Gaididei}}]{Sheka2015JPMT}%
  \BibitemOpen
  \bibfield  {author} {\bibinfo {author} {\bibfnamefont {D.~D.}\ \bibnamefont
  {Sheka}}, \bibinfo {author} {\bibfnamefont {V.~P.}\ \bibnamefont {Kravchuk}},
  \ and\ \bibinfo {author} {\bibfnamefont {Y.}~\bibnamefont {Gaididei}},\
  }\href {\doibase 10.1088/1751-8113/48/12/125202} {\bibfield  {journal}
  {\bibinfo  {journal} {J. Phys. A: Math. Theor.}\ }\textbf {\bibinfo {volume}
  {48}},\ \bibinfo {pages} {125202} (\bibinfo {year} {2015})}\BibitemShut
  {NoStop}%
\bibitem [{\citenamefont {Streubel}\ \emph {et~al.}(2016)\citenamefont
  {Streubel}, \citenamefont {Fischer}, \citenamefont {Kronast}, \citenamefont
  {Kravchuk}, \citenamefont {Sheka}, \citenamefont {Gaididei}, \citenamefont
  {Schmidt},\ and\ \citenamefont {Makarov}}]{Streubel2016}%
  \BibitemOpen
  \bibfield  {author} {\bibinfo {author} {\bibfnamefont {R.}~\bibnamefont
  {Streubel}}, \bibinfo {author} {\bibfnamefont {P.}~\bibnamefont {Fischer}},
  \bibinfo {author} {\bibfnamefont {F.}~\bibnamefont {Kronast}}, \bibinfo
  {author} {\bibfnamefont {V.~P.}\ \bibnamefont {Kravchuk}}, \bibinfo {author}
  {\bibfnamefont {D.~D.}\ \bibnamefont {Sheka}}, \bibinfo {author}
  {\bibfnamefont {Y.}~\bibnamefont {Gaididei}}, \bibinfo {author}
  {\bibfnamefont {O.~G.}\ \bibnamefont {Schmidt}}, \ and\ \bibinfo {author}
  {\bibfnamefont {D.}~\bibnamefont {Makarov}},\ }\href {\doibase
  10.1088/0022-3727/49/36/363001} {\bibfield  {journal} {\bibinfo  {journal}
  {J. Phys. D: Appl. Phys.}\ }\textbf {\bibinfo {volume} {49}},\ \bibinfo
  {pages} {363001} (\bibinfo {year} {2016})}\BibitemShut {NoStop}%
\bibitem [{\citenamefont {Ot{\'{a}}lora}\ \emph {et~al.}(2016)\citenamefont
  {Ot{\'{a}}lora}, \citenamefont {Yan}, \citenamefont {Schultheiss},
  \citenamefont {Hertel},\ and\ \citenamefont {K{\'{a}}kay}}]{Otalora2016}%
  \BibitemOpen
  \bibfield  {author} {\bibinfo {author} {\bibfnamefont {J.~A.}\ \bibnamefont
  {Ot{\'{a}}lora}}, \bibinfo {author} {\bibfnamefont {M.}~\bibnamefont {Yan}},
  \bibinfo {author} {\bibfnamefont {H.}~\bibnamefont {Schultheiss}}, \bibinfo
  {author} {\bibfnamefont {R.}~\bibnamefont {Hertel}}, \ and\ \bibinfo {author}
  {\bibfnamefont {A.}~\bibnamefont {K{\'{a}}kay}},\ }\href {\doibase
  10.1103/PhysRevLett.117.227203} {\bibfield  {journal} {\bibinfo  {journal}
  {Phys. Rev. Lett.}\ }\textbf {\bibinfo {volume} {117}},\ \bibinfo {pages}
  {227203} (\bibinfo {year} {2016})}\BibitemShut {NoStop}%
\bibitem [{\citenamefont {Tretiakov}\ \emph {et~al.}(2017)\citenamefont
  {Tretiakov}, \citenamefont {Morini}, \citenamefont {Vasylkevych},\ and\
  \citenamefont {Slastikov}}]{Tretiakov2017}%
  \BibitemOpen
  \bibfield  {author} {\bibinfo {author} {\bibfnamefont {O.~A.}\ \bibnamefont
  {Tretiakov}}, \bibinfo {author} {\bibfnamefont {M.}~\bibnamefont {Morini}},
  \bibinfo {author} {\bibfnamefont {S.}~\bibnamefont {Vasylkevych}}, \ and\
  \bibinfo {author} {\bibfnamefont {V.}~\bibnamefont {Slastikov}},\ }\href
  {\doibase 10.1103/PhysRevLett.119.077203} {\bibfield  {journal} {\bibinfo
  {journal} {Phys. Rev. Lett.}\ }\textbf {\bibinfo {volume} {119}},\ \bibinfo
  {pages} {077203} (\bibinfo {year} {2017})}\BibitemShut {NoStop}%
\bibitem [{\citenamefont {Charilaou}\ and\ \citenamefont
  {L{\"{o}}ffler}(2017)}]{Charilaou2017}%
  \BibitemOpen
  \bibfield  {author} {\bibinfo {author} {\bibfnamefont {M.}~\bibnamefont
  {Charilaou}}\ and\ \bibinfo {author} {\bibfnamefont {J.~F.}\ \bibnamefont
  {L{\"{o}}ffler}},\ }\href {\doibase 10.1103/PhysRevB.95.024409} {\bibfield
  {journal} {\bibinfo  {journal} {Phys. Rev. B}\ }\textbf {\bibinfo {volume}
  {95}},\ \bibinfo {pages} {024409} (\bibinfo {year} {2017})}\BibitemShut
  {NoStop}%
\bibitem [{\citenamefont {Moreno}\ \emph {et~al.}(2017)\citenamefont {Moreno},
  \citenamefont {Carvalho-Santos}, \citenamefont {Espejo}, \citenamefont
  {Laroze}, \citenamefont {Chubykalo-Fesenko},\ and\ \citenamefont
  {Altbir}}]{Moreno2017}%
  \BibitemOpen
  \bibfield  {author} {\bibinfo {author} {\bibfnamefont {R.}~\bibnamefont
  {Moreno}}, \bibinfo {author} {\bibfnamefont {V.~L.}\ \bibnamefont
  {Carvalho-Santos}}, \bibinfo {author} {\bibfnamefont {A.~P.}\ \bibnamefont
  {Espejo}}, \bibinfo {author} {\bibfnamefont {D.}~\bibnamefont {Laroze}},
  \bibinfo {author} {\bibfnamefont {O.}~\bibnamefont {Chubykalo-Fesenko}}, \
  and\ \bibinfo {author} {\bibfnamefont {D.}~\bibnamefont {Altbir}},\ }\href
  {\doibase 10.1103/PhysRevB.96.184401} {\bibfield  {journal} {\bibinfo
  {journal} {Phys. Rev. B}\ }\textbf {\bibinfo {volume} {96}},\ \bibinfo
  {pages} {184401} (\bibinfo {year} {2017})}\BibitemShut {NoStop}%
\bibitem [{\citenamefont {Ball}\ \emph {et~al.}(2017)\citenamefont {Ball},
  \citenamefont {G{\"{u}}nther}, \citenamefont {Fritzsche}, \citenamefont
  {Lenz}, \citenamefont {Varvaro}, \citenamefont {Laureti}, \citenamefont
  {Makarov}, \citenamefont {M{\"{u}}cklich}, \citenamefont {Facsko},
  \citenamefont {Albrecht},\ and\ \citenamefont {Fassbender}}]{Ball2017}%
  \BibitemOpen
  \bibfield  {author} {\bibinfo {author} {\bibfnamefont {D.~K.}\ \bibnamefont
  {Ball}}, \bibinfo {author} {\bibfnamefont {S.}~\bibnamefont {G{\"{u}}nther}},
  \bibinfo {author} {\bibfnamefont {M.}~\bibnamefont {Fritzsche}}, \bibinfo
  {author} {\bibfnamefont {K.}~\bibnamefont {Lenz}}, \bibinfo {author}
  {\bibfnamefont {G.}~\bibnamefont {Varvaro}}, \bibinfo {author} {\bibfnamefont
  {S.}~\bibnamefont {Laureti}}, \bibinfo {author} {\bibfnamefont
  {D.}~\bibnamefont {Makarov}}, \bibinfo {author} {\bibfnamefont
  {A.}~\bibnamefont {M{\"{u}}cklich}}, \bibinfo {author} {\bibfnamefont
  {S.}~\bibnamefont {Facsko}}, \bibinfo {author} {\bibfnamefont
  {M.}~\bibnamefont {Albrecht}}, \ and\ \bibinfo {author} {\bibfnamefont
  {J.}~\bibnamefont {Fassbender}},\ }\href {\doibase 10.1088/1361-6463/aa5c26}
  {\bibfield  {journal} {\bibinfo  {journal} {J. Phys. D}\ }\textbf {\bibinfo
  {volume} {50}},\ \bibinfo {pages} {115004} (\bibinfo {year}
  {2017})}\BibitemShut {NoStop}%
\bibitem [{\citenamefont {Vojkovic}\ \emph {et~al.}(2017)\citenamefont
  {Vojkovic}, \citenamefont {Carvalho-Santos}, \citenamefont {Fonseca},\ and\
  \citenamefont {Nunez}}]{Vojkovic2017}%
  \BibitemOpen
  \bibfield  {author} {\bibinfo {author} {\bibfnamefont {S.}~\bibnamefont
  {Vojkovic}}, \bibinfo {author} {\bibfnamefont {V.~L.}\ \bibnamefont
  {Carvalho-Santos}}, \bibinfo {author} {\bibfnamefont {J.~M.}\ \bibnamefont
  {Fonseca}}, \ and\ \bibinfo {author} {\bibfnamefont {A.~S.}\ \bibnamefont
  {Nunez}},\ }\href {\doibase 10.1063/1.4977983} {\bibfield  {journal}
  {\bibinfo  {journal} {J. Appl. Phys.}\ }\textbf {\bibinfo {volume} {121}},\
  \bibinfo {pages} {113906} (\bibinfo {year} {2017})}\BibitemShut {NoStop}%
\bibitem [{\citenamefont {Zvyagin}(2017)}]{Zvyagin2017}%
  \BibitemOpen
  \bibfield  {author} {\bibinfo {author} {\bibfnamefont {A.~A.}\ \bibnamefont
  {Zvyagin}},\ }\href {\doibase 10.1103/PhysRevB.95.165141} {\bibfield
  {journal} {\bibinfo  {journal} {Phys. Rev. B}\ }\textbf {\bibinfo {volume}
  {95}},\ \bibinfo {pages} {165141} (\bibinfo {year} {2017})}\BibitemShut
  {NoStop}%
\bibitem [{\citenamefont {Kondo}(2005)}]{Kondo2005}%
  \BibitemOpen
  \bibfield  {author} {\bibinfo {author} {\bibfnamefont {J.}~\bibnamefont
  {Kondo}},\ }\href {\doibase 10.1143/JPSJ.74.1} {\bibfield  {journal}
  {\bibinfo  {journal} {J. Phys. Soc. Jpn.}\ }\textbf {\bibinfo {volume}
  {74}},\ \bibinfo {pages} {1} (\bibinfo {year} {2005})}\BibitemShut {NoStop}%
\bibitem [{\citenamefont {Maeno}\ \emph {et~al.}(2012)\citenamefont {Maeno},
  \citenamefont {Kittaka}, \citenamefont {Nomura}, \citenamefont {Yonezawa},\
  and\ \citenamefont {Ishida}}]{Maeno2012}%
  \BibitemOpen
  \bibfield  {author} {\bibinfo {author} {\bibfnamefont {Y.}~\bibnamefont
  {Maeno}}, \bibinfo {author} {\bibfnamefont {S.}~\bibnamefont {Kittaka}},
  \bibinfo {author} {\bibfnamefont {T.}~\bibnamefont {Nomura}}, \bibinfo
  {author} {\bibfnamefont {S.}~\bibnamefont {Yonezawa}}, \ and\ \bibinfo
  {author} {\bibfnamefont {K.}~\bibnamefont {Ishida}},\ }\href {\doibase
  10.1143/JPSJ.81.011009} {\bibfield  {journal} {\bibinfo  {journal} {J. Phys.
  Soc. Jpn.}\ }\textbf {\bibinfo {volume} {81}},\ \bibinfo {pages} {011009}
  (\bibinfo {year} {2012})}\BibitemShut {NoStop}%
\bibitem [{\citenamefont {Shekhter}\ \emph {et~al.}(2016)\citenamefont
  {Shekhter}, \citenamefont {Jonson},\ and\ \citenamefont
  {Aharony}}]{Shekhter2016}%
  \BibitemOpen
  \bibfield  {author} {\bibinfo {author} {\bibfnamefont {R.~I.}\ \bibnamefont
  {Shekhter}}, \bibinfo {author} {\bibfnamefont {M.}~\bibnamefont {Jonson}}, \
  and\ \bibinfo {author} {\bibfnamefont {A.}~\bibnamefont {Aharony}},\ }\href
  {\doibase 10.1103/PhysRevLett.116.217001} {\bibfield  {journal} {\bibinfo
  {journal} {Phys. Rev. Lett.}\ }\textbf {\bibinfo {volume} {116}},\ \bibinfo
  {pages} {217001} (\bibinfo {year} {2016})}\BibitemShut {NoStop}%
\bibitem [{\citenamefont {Can}\ \emph {et~al.}(2014)\citenamefont {Can},
  \citenamefont {Laskin},\ and\ \citenamefont {Wiegmann}}]{Can2014}%
  \BibitemOpen
  \bibfield  {author} {\bibinfo {author} {\bibfnamefont {T.}~\bibnamefont
  {Can}}, \bibinfo {author} {\bibfnamefont {M.}~\bibnamefont {Laskin}}, \ and\
  \bibinfo {author} {\bibfnamefont {P.}~\bibnamefont {Wiegmann}},\ }\href
  {\doibase 10.1103/PhysRevLett.113.046803} {\bibfield  {journal} {\bibinfo
  {journal} {Phys. Rev. Lett.}\ }\textbf {\bibinfo {volume} {113}},\ \bibinfo
  {pages} {046803} (\bibinfo {year} {2014})}\BibitemShut {NoStop}%
\bibitem [{\citenamefont {Can}\ \emph {et~al.}(2016)\citenamefont {Can},
  \citenamefont {Chiu}, \citenamefont {Laskin},\ and\ \citenamefont
  {Wiegmann}}]{Can2016}%
  \BibitemOpen
  \bibfield  {author} {\bibinfo {author} {\bibfnamefont {T.}~\bibnamefont
  {Can}}, \bibinfo {author} {\bibfnamefont {Y.~H.}\ \bibnamefont {Chiu}},
  \bibinfo {author} {\bibfnamefont {M.}~\bibnamefont {Laskin}}, \ and\ \bibinfo
  {author} {\bibfnamefont {P.}~\bibnamefont {Wiegmann}},\ }\href {\doibase
  10.1103/PhysRevLett.117.266803} {\bibfield  {journal} {\bibinfo  {journal}
  {Phys. Rev. Lett.}\ }\textbf {\bibinfo {volume} {117}},\ \bibinfo {pages}
  {266803} (\bibinfo {year} {2016})}\BibitemShut {NoStop}%
\bibitem [{\citenamefont {Szameit}\ \emph {et~al.}(2010)\citenamefont
  {Szameit}, \citenamefont {Dreisow}, \citenamefont {Heinrich}, \citenamefont
  {Keil}, \citenamefont {Nolte}, \citenamefont {T{\"{u}}nnermann},\ and\
  \citenamefont {Longhi}}]{Szameit2010}%
  \BibitemOpen
  \bibfield  {author} {\bibinfo {author} {\bibfnamefont {A.}~\bibnamefont
  {Szameit}}, \bibinfo {author} {\bibfnamefont {F.}~\bibnamefont {Dreisow}},
  \bibinfo {author} {\bibfnamefont {M.}~\bibnamefont {Heinrich}}, \bibinfo
  {author} {\bibfnamefont {R.}~\bibnamefont {Keil}}, \bibinfo {author}
  {\bibfnamefont {S.}~\bibnamefont {Nolte}}, \bibinfo {author} {\bibfnamefont
  {A.}~\bibnamefont {T{\"{u}}nnermann}}, \ and\ \bibinfo {author}
  {\bibfnamefont {S.}~\bibnamefont {Longhi}},\ }\href {\doibase
  10.1103/PhysRevLett.104.150403} {\bibfield  {journal} {\bibinfo  {journal}
  {Phys. Rev. Lett.}\ }\textbf {\bibinfo {volume} {104}},\ \bibinfo {pages}
  {150403} (\bibinfo {year} {2010})}\BibitemShut {NoStop}%
\bibitem [{\citenamefont {Schultheiss}\ \emph {et~al.}(2010)\citenamefont
  {Schultheiss}, \citenamefont {Batz}, \citenamefont {Szameit}, \citenamefont
  {Dreisow}, \citenamefont {Nolte}, \citenamefont {T{\"{u}}nnermann},
  \citenamefont {Longhi},\ and\ \citenamefont {Peschel}}]{Schultheiss2010}%
  \BibitemOpen
  \bibfield  {author} {\bibinfo {author} {\bibfnamefont {V.~H.}\ \bibnamefont
  {Schultheiss}}, \bibinfo {author} {\bibfnamefont {S.}~\bibnamefont {Batz}},
  \bibinfo {author} {\bibfnamefont {A.}~\bibnamefont {Szameit}}, \bibinfo
  {author} {\bibfnamefont {F.}~\bibnamefont {Dreisow}}, \bibinfo {author}
  {\bibfnamefont {S.}~\bibnamefont {Nolte}}, \bibinfo {author} {\bibfnamefont
  {A.}~\bibnamefont {T{\"{u}}nnermann}}, \bibinfo {author} {\bibfnamefont
  {S.}~\bibnamefont {Longhi}}, \ and\ \bibinfo {author} {\bibfnamefont
  {U.}~\bibnamefont {Peschel}},\ }\href {\doibase
  10.1103/PhysRevLett.105.143901} {\bibfield  {journal} {\bibinfo  {journal}
  {Phys. Rev. Lett.}\ }\textbf {\bibinfo {volume} {105}},\ \bibinfo {pages}
  {143901} (\bibinfo {year} {2010})}\BibitemShut {NoStop}%
\bibitem [{\citenamefont {Nagasawa}\ \emph {et~al.}(2012)\citenamefont
  {Nagasawa}, \citenamefont {Takagi}, \citenamefont {Kunihashi}, \citenamefont
  {Kohda},\ and\ \citenamefont {Nitta}}]{Nagasawa2012}%
  \BibitemOpen
  \bibfield  {author} {\bibinfo {author} {\bibfnamefont {F.}~\bibnamefont
  {Nagasawa}}, \bibinfo {author} {\bibfnamefont {J.}~\bibnamefont {Takagi}},
  \bibinfo {author} {\bibfnamefont {Y.}~\bibnamefont {Kunihashi}}, \bibinfo
  {author} {\bibfnamefont {M.}~\bibnamefont {Kohda}}, \ and\ \bibinfo {author}
  {\bibfnamefont {J.}~\bibnamefont {Nitta}},\ }\href {\doibase
  10.1103/PhysRevLett.108.086801} {\bibfield  {journal} {\bibinfo  {journal}
  {Phys. Rev. Lett.}\ }\textbf {\bibinfo {volume} {108}},\ \bibinfo {pages}
  {086801} (\bibinfo {year} {2012})}\BibitemShut {NoStop}%
\bibitem [{\citenamefont {Nagasawa}\ \emph {et~al.}(2013)\citenamefont
  {Nagasawa}, \citenamefont {Frustaglia}, \citenamefont {Saarikoski},
  \citenamefont {Richter},\ and\ \citenamefont {Nitta}}]{Nagasawa2013}%
  \BibitemOpen
  \bibfield  {author} {\bibinfo {author} {\bibfnamefont {F.}~\bibnamefont
  {Nagasawa}}, \bibinfo {author} {\bibfnamefont {D.}~\bibnamefont
  {Frustaglia}}, \bibinfo {author} {\bibfnamefont {H.}~\bibnamefont
  {Saarikoski}}, \bibinfo {author} {\bibfnamefont {K.}~\bibnamefont {Richter}},
  \ and\ \bibinfo {author} {\bibfnamefont {J.}~\bibnamefont {Nitta}},\ }\href
  {\doibase 10.1038/ncomms3526} {\bibfield  {journal} {\bibinfo  {journal}
  {Nat. Commun.}\ }\textbf {\bibinfo {volume} {4}},\ \bibinfo {pages} {2526}
  (\bibinfo {year} {2013})}\BibitemShut {NoStop}%
\bibitem [{\citenamefont {Saarikoski}\ \emph {et~al.}(2015)\citenamefont
  {Saarikoski}, \citenamefont {Vazquez-Lozano}, \citenamefont {Baltanas},
  \citenamefont {Nagasawa}, \citenamefont {Nitta},\ and\ \citenamefont
  {Frustaglia}}]{Saarikoski2015}%
  \BibitemOpen
  \bibfield  {author} {\bibinfo {author} {\bibfnamefont {H.}~\bibnamefont
  {Saarikoski}}, \bibinfo {author} {\bibfnamefont {J.~E.}\ \bibnamefont
  {Vazquez-Lozano}}, \bibinfo {author} {\bibfnamefont {J.~P.}\ \bibnamefont
  {Baltanas}}, \bibinfo {author} {\bibfnamefont {F.}~\bibnamefont {Nagasawa}},
  \bibinfo {author} {\bibfnamefont {J.}~\bibnamefont {Nitta}}, \ and\ \bibinfo
  {author} {\bibfnamefont {D.}~\bibnamefont {Frustaglia}},\ }\href {\doibase
  10.1103/PhysRevB.91.241406} {\bibfield  {journal} {\bibinfo  {journal} {Phys.
  Rev. B}\ }\textbf {\bibinfo {volume} {91}},\ \bibinfo {pages} {241406(R)}
  (\bibinfo {year} {2015})}\BibitemShut {NoStop}%
\bibitem [{\citenamefont {Saarikoski}\ \emph {et~al.}(2016)\citenamefont
  {Saarikoski}, \citenamefont {Baltan{\'{a}}s}, \citenamefont
  {V{\'{a}}zquez-Lozano}, \citenamefont {Nitta},\ and\ \citenamefont
  {Frustaglia}}]{Saarikoski2016}%
  \BibitemOpen
  \bibfield  {author} {\bibinfo {author} {\bibfnamefont {H.}~\bibnamefont
  {Saarikoski}}, \bibinfo {author} {\bibfnamefont {J.~P.}\ \bibnamefont
  {Baltan{\'{a}}s}}, \bibinfo {author} {\bibfnamefont {J.~E.}\ \bibnamefont
  {V{\'{a}}zquez-Lozano}}, \bibinfo {author} {\bibfnamefont {J.}~\bibnamefont
  {Nitta}}, \ and\ \bibinfo {author} {\bibfnamefont {D.}~\bibnamefont
  {Frustaglia}},\ }\href {\doibase 10.1088/0953-8984/28/16/166002} {\bibfield
  {journal} {\bibinfo  {journal} {J. Phys.: Condens. Matter}\ }\textbf
  {\bibinfo {volume} {28}},\ \bibinfo {pages} {166002} (\bibinfo {year}
  {2016})}\BibitemShut {NoStop}%
\bibitem [{\citenamefont {Avishai}\ and\ \citenamefont
  {Band}(2017)}]{Avishai2017}%
  \BibitemOpen
  \bibfield  {author} {\bibinfo {author} {\bibfnamefont {Y.}~\bibnamefont
  {Avishai}}\ and\ \bibinfo {author} {\bibfnamefont {Y.~B.}\ \bibnamefont
  {Band}},\ }\href {\doibase 10.1103/PhysRevB.95.104429} {\bibfield  {journal}
  {\bibinfo  {journal} {Phys. Rev. B}\ }\textbf {\bibinfo {volume} {95}},\
  \bibinfo {pages} {104429} (\bibinfo {year} {2017})}\BibitemShut {NoStop}%
\bibitem [{\citenamefont {Hamada}\ \emph {et~al.}(2015)\citenamefont {Hamada},
  \citenamefont {Yokoyama},\ and\ \citenamefont {Murakami}}]{Hamada2015}%
  \BibitemOpen
  \bibfield  {author} {\bibinfo {author} {\bibfnamefont {M.}~\bibnamefont
  {Hamada}}, \bibinfo {author} {\bibfnamefont {T.}~\bibnamefont {Yokoyama}}, \
  and\ \bibinfo {author} {\bibfnamefont {S.}~\bibnamefont {Murakami}},\ }\href
  {\doibase 10.1103/PhysRevB.92.060409} {\bibfield  {journal} {\bibinfo
  {journal} {Phys. Rev. B}\ }\textbf {\bibinfo {volume} {92}},\ \bibinfo
  {pages} {060409(R)} (\bibinfo {year} {2015})}\BibitemShut {NoStop}%
\bibitem [{\citenamefont {Dong}\ and\ \citenamefont {Niu}()}]{Dong2018}%
  \BibitemOpen
  \bibfield  {author} {\bibinfo {author} {\bibfnamefont {L.}~\bibnamefont
  {Dong}}\ and\ \bibinfo {author} {\bibfnamefont {Q.}~\bibnamefont {Niu}},\
  }\href {http://arxiv.org/abs/1802.02887} {\ \bibinfo {pages}
  {arXiv:1802.02887}}\BibitemShut {NoStop}%
\bibitem [{\citenamefont {Brendel}\ \emph {et~al.}(2017)\citenamefont
  {Brendel}, \citenamefont {Peano}, \citenamefont {Painter},\ and\
  \citenamefont {Marquardt}}]{Brendel2017}%
  \BibitemOpen
  \bibfield  {author} {\bibinfo {author} {\bibfnamefont {C.}~\bibnamefont
  {Brendel}}, \bibinfo {author} {\bibfnamefont {V.}~\bibnamefont {Peano}},
  \bibinfo {author} {\bibfnamefont {O.}~\bibnamefont {Painter}}, \ and\
  \bibinfo {author} {\bibfnamefont {F.}~\bibnamefont {Marquardt}},\ }\href
  {\doibase 10.1073/pnas.1615503114} {\bibfield  {journal} {\bibinfo  {journal}
  {Proc. Natl. Acad. Sci. U.S.A.}\ }\textbf {\bibinfo {volume} {114}},\
  \bibinfo {pages} {E3390} (\bibinfo {year} {2017})}\BibitemShut {NoStop}%
\bibitem [{\citenamefont {G{\"{o}}hler}\ \emph {et~al.}(2011)\citenamefont
  {G{\"{o}}hler}, \citenamefont {Hamelbeck}, \citenamefont {Markus},
  \citenamefont {Kettner}, \citenamefont {Hanne}, \citenamefont {Vager},
  \citenamefont {Naaman},\ and\ \citenamefont {Zacharias}}]{Gohler2011}%
  \BibitemOpen
  \bibfield  {author} {\bibinfo {author} {\bibfnamefont {B.}~\bibnamefont
  {G{\"{o}}hler}}, \bibinfo {author} {\bibfnamefont {V.}~\bibnamefont
  {Hamelbeck}}, \bibinfo {author} {\bibfnamefont {T.~Z.}\ \bibnamefont
  {Markus}}, \bibinfo {author} {\bibfnamefont {M.}~\bibnamefont {Kettner}},
  \bibinfo {author} {\bibfnamefont {G.~F.}\ \bibnamefont {Hanne}}, \bibinfo
  {author} {\bibfnamefont {Z.}~\bibnamefont {Vager}}, \bibinfo {author}
  {\bibfnamefont {R.}~\bibnamefont {Naaman}}, \ and\ \bibinfo {author}
  {\bibfnamefont {H.}~\bibnamefont {Zacharias}},\ }\href {\doibase
  10.1126/science.1199339} {\bibfield  {journal} {\bibinfo  {journal}
  {Science}\ }\textbf {\bibinfo {volume} {331}},\ \bibinfo {pages} {894}
  (\bibinfo {year} {2011})}\BibitemShut {NoStop}%
\bibitem [{\citenamefont {Guo}\ and\ \citenamefont {Sun}(2012)}]{Guo2012}%
  \BibitemOpen
  \bibfield  {author} {\bibinfo {author} {\bibfnamefont {A.~M.}\ \bibnamefont
  {Guo}}\ and\ \bibinfo {author} {\bibfnamefont {Q.~F.}\ \bibnamefont {Sun}},\
  }\href {\doibase 10.1103/PhysRevLett.108.218102} {\bibfield  {journal}
  {\bibinfo  {journal} {Phys. Rev. Lett.}\ }\textbf {\bibinfo {volume} {108}},\
  \bibinfo {pages} {218102} (\bibinfo {year} {2012})}\BibitemShut {NoStop}%
\bibitem [{\citenamefont {Gutierrez}\ \emph {et~al.}(2013)\citenamefont
  {Gutierrez}, \citenamefont {D{\'{i}}az}, \citenamefont {Gaul}, \citenamefont
  {Brumme}, \citenamefont {Dom{\'{i}}nguez-Adame},\ and\ \citenamefont
  {Cuniberti}}]{Gutierrez2013}%
  \BibitemOpen
  \bibfield  {author} {\bibinfo {author} {\bibfnamefont {R.}~\bibnamefont
  {Gutierrez}}, \bibinfo {author} {\bibfnamefont {E.}~\bibnamefont
  {D{\'{i}}az}}, \bibinfo {author} {\bibfnamefont {C.}~\bibnamefont {Gaul}},
  \bibinfo {author} {\bibfnamefont {T.}~\bibnamefont {Brumme}}, \bibinfo
  {author} {\bibfnamefont {F.}~\bibnamefont {Dom{\'{i}}nguez-Adame}}, \ and\
  \bibinfo {author} {\bibfnamefont {G.}~\bibnamefont {Cuniberti}},\ }\href
  {\doibase 10.1021/jp401705x} {\bibfield  {journal} {\bibinfo  {journal} {J.
  Phys. Chem. C}\ }\textbf {\bibinfo {volume} {117}},\ \bibinfo {pages} {22276}
  (\bibinfo {year} {2013})}\BibitemShut {NoStop}%
\bibitem [{\citenamefont {Strange}(1998)}]{book_Strange}%
  \BibitemOpen
  \bibfield  {author} {\bibinfo {author} {\bibfnamefont {P.}~\bibnamefont
  {Strange}},\ }\href@noop {} {\emph {\bibinfo {title} {{Relativistic Quantum
  Mechanics: With Applications in Condensed Matter and Atomic Physics}}}}\
  (\bibinfo  {publisher} {Cambridge University Press},\ \bibinfo {year}
  {1998})\BibitemShut {NoStop}%
\bibitem [{\citenamefont {Nishijima}(1973)}]{book_Nishijima}%
  \BibitemOpen
  \bibfield  {author} {\bibinfo {author} {\bibfnamefont {K.}~\bibnamefont
  {Nishijima}},\ }\href@noop {} {\emph {\bibinfo {title} {{Relativistic Quantum
  Mechanics}}}}\ (\bibinfo  {publisher} {Baifukan, in Japanese},\ \bibinfo
  {year} {1973})\BibitemShut {NoStop}%
\end{thebibliography}%

\end{document}